\documentclass[conf]{new-aiaa} 
\usepackage[utf8]{inputenc}
\usepackage{textcomp}

\usepackage{graphicx}
\usepackage{amsmath}
\usepackage[version=4]{mhchem}
\usepackage{siunitx}
\usepackage{longtable,tabularx}
\usepackage{amsthm,amssymb}
\usepackage{hyperref}
\usepackage{subcaption}
\usepackage{xcolor}
\usepackage[section]{placeins} 

\newcommand{\bq}{\mathbf{q}}
\newcommand{\bp}{\mathbf{p}}
\newcommand{\bu}{\mathbf{u}}
\newcommand{\bv}{\mathbf{v}}
\newcommand{\bhc}{\hat{\mathbf{c}}}
\newcommand{\bA}{\mathbf{A}}
\newcommand{\bB}{\mathbf{B}}
\newcommand{\bH}{\mathbf{H}}
\newcommand{\bF}{\mathbf{F}}
\newcommand{\bV}{\mathbf{V}}
\newcommand{\bQ}{\mathbf{Q}}

\newcommand{\CHfour}{c_1}
\newcommand{\Otwo}{c_2}

\newcommand{\Rev}[1]{\color{black} {#1}}

\title{Learning physics-based reduced-order models \\for a single-injector combustion process}
\author[1]{Renee Swischuk\footnote{Graduate Student, MIT Center for Computational Engineering, swischuk@mit.edu,  Student Member AIAA.}}
\author[2]{Boris Kramer\footnote{Assistant Professor, Department of Mechanical and Aerospace Engineering, bmkramer@ucsd.edu, Member AIAA}}
\author[3]{Cheng Huang\footnote{Assistant Research Scientist, Department of Aerospace Engineering, huangche@umich.edu, Member AIAA}}
\author[4]{Karen Willcox\footnote{Director, Oden Institute for Computational Engineering and Sciences, kwillcox@oden.utexas.edu, AIAA Fellow.}}
\affil[1]{Massachusetts Institute of Technology, Cambridge, MA, 02139}
\affil[2]{University of California, San Diego, CA, 92122}
\affil[3]{University of Michigan, Ann Arbor, MI, 48109}
\affil[4]{University of Texas at Austin, Austin, TX, 78712}

\begin{document}

\maketitle

\begin{abstract}
This paper presents a physics-based data-driven method to learn predictive reduced-order models (ROMs) from high-fidelity simulations, and illustrates it in the challenging context of a single-injector combustion process.
The method combines the perspectives of model reduction and machine learning.
Model reduction brings in the physics of the problem, constraining the ROM predictions to lie on a subspace defined by the governing equations. This is achieved by defining the ROM in proper orthogonal decomposition (POD) coordinates, which embed the rich physics information contained in solution snapshots of a high-fidelity computational fluid dynamics (CFD) model.
The machine learning perspective brings the flexibility to use transformed physical variables to define the POD basis. This is in contrast to traditional model reduction approaches that are constrained to use the physical variables of the high-fidelity code.
Combining the two perspectives, the approach identifies a set of transformed physical variables that expose quadratic structure in the combustion governing equations and learns a quadratic ROM from transformed snapshot data. This learning does not require access to the high-fidelity model implementation.
Numerical experiments show that the ROM accurately predicts temperature, pressure, velocity, species concentrations, and the limit-cycle amplitude, with speedups of more than five orders of magnitude over high-fidelity models. 
{\Rev{Our ROM simulation is shown to be predictive 200\% past the training interval. }}
ROM-predicted pressure traces accurately match the phase of the pressure signal and yield good approximations of the limit-cycle amplitude.
\end{abstract}

\section*{Nomenclature}

{\renewcommand\arraystretch{1.0}
	\noindent
	\begin{longtable*}{@{}l @{\quad=\quad} l@{}}
		$\bA$ & System matrix for linear part \\
		$\bB$ & Input matrix\\
		$\bH$ & Matricized quadratic tensor\\
		$\bq(t)$ & State vector in finite dimensions\\			
		$\bu(t)$ & External input vector\\		
		$\bQ$ & Snapshot matrix\\		
		$\bV$ & Matrix of POD basis vectors\\
		$x,y$ & Spatial coordinates \\	
		$d$ & Number of physical variables\\
		$n_x$ & Spatial discretization dimension\\
		$r$ & Reduced model dimension \\
		$t$ & Time \\				
		$\vec{q}_p(t,x,y), \ \vec{q}_c(t,x,y), \ \vec{q}_L(t,x,y)$ & State vector in primitive, conservative, and learning variables \\	
		$p(t,x,y), \ \bp(t)$ & Pressure, continuous and discretized\\
		$T(t,x,y), \ \mathbf{T}(t)$ & Temperature, continuous and discretized\\
		$\rho(t,x,y), \ \boldsymbol{\rho}(t)$ & Density, continuous and discretized\\	
		$\xi(t,x,y), \ \boldsymbol{\xi}(t)$ & Specific volume, continuous and discretized\\
		$v_x(t,x,y), \ \bv_x(t)$ & Velocity in $x$ direction, continuous and discretized\\	
		$Y_l(t,x,y)$ & Species mass fraction, $l=1,2, \ldots, n_{\text{sp}}$\\	
		$c_l(t,x,y)$ & Species molar concentrations, $l=1,2, \ldots, n_{\text{sp}}$, also denoted as $[S]$\\			

		$\otimes$ & Kronecker product \\
		$\widehat{\cdot}$& Notation for ROM quantities \\
		\multicolumn{2}{@{}l}{\textbf{Abbreviations}}\\
		CFD & Computational Fluid Dynamics \\	
		GEMS & General equation and mesh solver; a CFD code \\			
		PDE & Partial differential equation \\
		POD & Proper orthogonal decomposition \\
		ROM & Reduced-order model
	\end{longtable*}}

\section{Introduction}
This paper presents an approach to  learning low-dimensional  surrogate models for a complex, nonlinear, multi-physics, multi-scale {\Rev dynamical system} in form of a multi-species combustion process.
The need for repeated model evaluations in optimization, design, uncertainty quantification and control of aerospace systems has driven the development of reduced-order models (ROMs) for applications in aerodynamics~\cite{bui2004aerodynamic,tadmor2007low,Lieu2007,amsallem2010towards,berger2013reduced,brunton_rowley_williams_2013,KGBNB17SensingDMDbifurcationsFlows}, reacting flows~\cite{nguyen2014model,nguyen2017model,buffoni10MORreactingFlows} and combustion~\cite{cheng,huang2018exploration,hesthaven2019ROMCVRC}. ROMs combine the rich information embedded in high-fidelity simulations with the efficiency of low-dimensional surrogate models; yet, effective and robust ROM methods for nonlinear, multi-scale applications such as combustion have remained an open challenge. 

Most existing nonlinear model reduction methods are intrusive---that is, they derive the ROM by projecting the high-fidelity model operators onto a low-dimensional subspace. In doing so, the physics of the problem is embedded in the reduced-order representation. The proper orthogonal decomposition (POD)~\cite{lumley1967structure,sirovich87turbulence} is the most common way to define the low-dimensional subspace, using the singular value decomposition to identify low-dimensional structure based on training data.
For some problems, the projection approach is amenable to rigorous error analysis and structure-preservation guarantees \cite{veroy2002posteriori,grepl2007efficient, troltzsch2009pod,hesthaven2016certified}, but these rigorous guarantees do not apply to nonlinear, multi-physics, multi-scale models, for which projection-based ROMs remain challenging to implement (due to the need for access to the high-dimensional operators).
{\Rev The compressible flow setting of the combustion process poses numerous problems with respect to stability of the projection-based ROMs, see~\cite{rowley2004model,barone2009stable,serre2012reliable,kalashnikova2011stable,balajewicz2016minimal,carlberg2017galerkin} for several approaches to address this stability problem}. 
Furthermore, ROMs for these problems typically require relatively high dimensionality (and thus high cost) to avoid problems with robustness and stability \cite{huang2018exploration,huang2018challenges}. 
{\Rev For instance, Huang et al.~\cite{cheng} construct two separate ROMs for the same single-injector combustion simulation as presented herein, one that uses POD and the other uses least-squares Petrov-Galerkin projection. That work finds that well over 100 modes are necessary to obtain stable ROMs and sufficient accuracy. }

There is increasing attention to non-intrusive model reduction methods (sometimes called black-box or data-driven methods) that learn a model based on training data, without requiring explicit access to the high-fidelity model operators. The non-intrusive philosophy aligns directly with the field of machine learning, where representations such as neural networks have been shown to induce nonlinear model forms that can approximate many physical processes~\cite{hornik1989multilayer}. However, neural networks require a large amount of training data, limiting their utility when the data comes from expensive large-scale partial differential equation (PDE) simulations~\cite{swischuk2018physics}. Moreover, the parametrization of the learned low-dimensional model is critical to the predictive accuracy and success of the learned model---in particular, it is critical to determining whether the ROM can issue reliable predictions in regimes outside of the training data.

For large-scale PDE models, an important class of non-intrusive learning approaches tackle this challenge of model parametrization by embedding the structure of the problem into the learning formulation. Some approaches use sparse learning techniques to identify PDE model terms that explain the data~\cite{brunton2016discovering,rudy2017data,schaeffer2018extracting}.
{\Rev{Dynamic mode decomposition~\cite{schmid2010dynamic,rowley2009spectral} extracts spectral information of the infinite dimensional linear Koopman operator from observed data of the nonlinear system. This spectral information can then be used to build data-driven predictive models.}}
When the model can be expressed in the form of a dynamical system with polynomial terms, then the learning problem can be formulated as a parameter estimation problem, as in the operator inference approach of~\cite{Peherstorfer16DataDriven}. An important advantage of non-intrusive learning approaches is that the user has the flexibility to choose the variables that drive the learning. This opens the way for variable transformations that expose system structure and, in doing so, transform the ROM learning task into a structured form. In some cases, the governing PDEs naturally admit variable transformations that reveal polynomial form, such as the specific volume representation of the Euler equations~\cite{elizabeth}.
More generally, one can introduce new auxiliary variables to the problem---known as \textit{lifting}---to produce a system that is polynomial in its expanded set of state variables~\cite{gu2011,KW18nonlinearMORliftingPOD,KW_BT_liftedSystems_2019}. This allows for a much broader class of nonlinear systems to be learned using the operator inference framework.

In this work, we build on the operator inference framework to learn structured, polynomial ROMs from simulated snapshot data of a single-injector combustion model. While in our case the polynomial model parametrization is a model approximation, we show that the predictive capabilities of the learned ROM are excellent beyond the training data.
Our proposed approach  follows the steps below, which we develop in detail in the ensuing sections:
\begin{enumerate}
	\item We obtain high-dimensional simulation snapshot data for a spatially two-dimensional combustion process from the General Equation and Mesh Solver (GEMS) CFD code~\cite{GEMScode} developed at Purdue University. The governing equations and combustion problem setup are described in Section~\ref{sec:combModel}.
	\item We identify a set of state variables in which many of the terms in the governing equations have quadratic form. We 
	transform the snapshot data to these new state variables, as described in Section~\ref{sec:transform}.
	\item We use operator inference to learn a ROM that evolves the combustion dynamics in a low-dimensional subspace. Details of the model learning are given in Section~\ref{sec:rom} and Section~\ref{sec:implementation}.
\end{enumerate}
We present numerical results comparing our learned ROMs with GEMS test data in Section~\ref{sec:numerics} and conclude the paper in Section~\ref{sec:conclusion}.

\section{Combustion model} \label{sec:combModel}
Section~\ref{sec:compDomain} defines the computational domain under consideration, Section~\ref{sec:governing} presents the governing equations for the combustion model, and Section~\ref{sec:GEMS_discretization} briefly summarizes the numerical implementation. The combustion model follows the implementation of the General Equation and Mesh Solver (GEMS) CFD code \cite{GEMScode} and more details can be found in \cite{harvazinski2012modeling}. {\Rev The GEMS code has been successfully used for rocket engine simulations~\cite{cheng} and in high-pressure gas turbines~\cite{huang2019combustionGasTurbine}. }

\subsection{Computational domain} \label{sec:compDomain}
\begin{figure}[h!]
	\centering
	\begin{subfigure}[t]{0.47\textwidth}
		\includegraphics[width=\textwidth,trim={0 0 .1cm 0.1cm},clip]{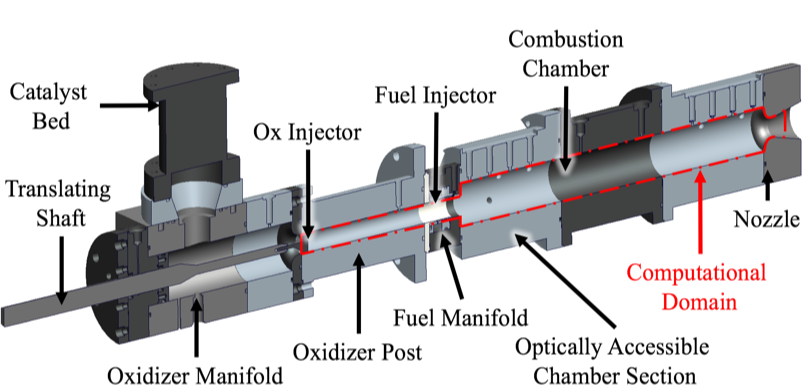}
		\caption{\label{fig:combustor1} Combustor assembly.}
	\end{subfigure}		
	\hfill
	\begin{subfigure}[t]{0.49\textwidth}
		\includegraphics[width=\textwidth]{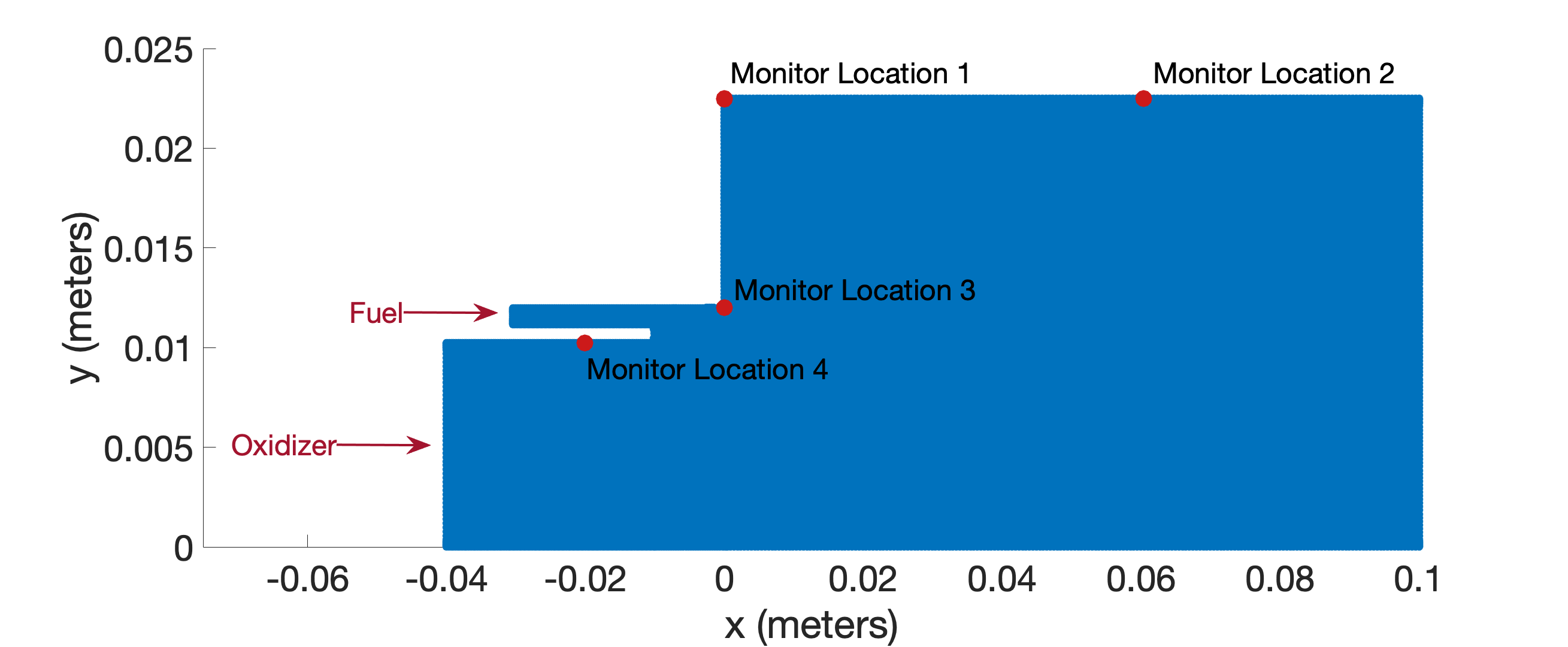}
		\caption{\label{fig:combustor2}  The upper half (due to symmetry) of the computational domain and the monitor locations where we measure the state variables.}	
	\end{subfigure}	
	\caption{Setup and geometry of single-injector combustor.}
\end{figure}

A single-injector combustor as in~\cite{domain} is shown in Figure~\ref{fig:combustor1}, with the computational domain outlined in red dashed lines. Our domain is a simplified two-dimensional version of the computational domain, shown in Figure~\ref{fig:combustor2}, which also shows the four locations where we monitor the state variables.

\subsection{Governing equations} \label{sec:governing}
The dynamics of the combustor are governed by the conservation equations for mass, momentum, energy and species mass fractions. For this two-dimensional problem, the conservation equations are
\begin{equation}
\frac{\partial \vec{q}_c}{\partial t} + \nabla \cdot (\vec{K} - \vec{K}_v ) = \vec{S} \label{eq:ns_conservative}
\end{equation}
and they describe the evolution of the conservative variables
\begin{equation*}
	\vec{q}_c =  [\rho \ \  \rho v_x \ \ \rho v_y \ \ \rho e \ \ \rho Y_1 \ \ \dots \ \ \rho Y_{n_{\textrm{sp}}} ]^{\top},
\end{equation*}
where $\rho$ is the density ($\frac{\text{kg}}{\text{m}^3}$), $v_x$ and $v_y$ are the $x$ and $y$ velocity ($\frac{\text{m}}{\text{s}}$), $e$ is the total energy ($\frac{\text{J}}{\text{m}^3}$), and $Y_l$ is the $l$th species mass fraction with $l = 1, 2, \ldots, n_{\textrm{sp}}$ and $n_{\textrm{sp}}$ is defined as the number of chemical species that are included in the model.

The total energy is defined as
\begin{equation}\label{eq:e}
e =  \sum_{l=1}^{n_{\textrm{sp}}} h_l Y_l + \frac{1}{2}\left(v^2_x+v^2_y\right) - \frac{p}{\rho} = h^0 - \frac{p}{\rho} ,
\end{equation}
where pressure $p$ is given in (Pa),  $h_l = h_l(T)$ is the enthalpy corresponding to the $l$th species and is a highly nonlinear function of temperature, $T$, and $h^0$ is the stagnation enthalpy.
The inviscid flux $\vec{K}$ and viscous flux  $\vec{K}_v$ in Eq.~\eqref{eq:ns_conservative} are
\begin{equation*}
\vec{K} =
\begin{bmatrix} \rho v_x \\ \rho v_x^2 + p \\ \rho v_x v_y \\ \rho v_x e + p v_x \\ \rho v_x Y_1 \\ \vdots \\ \rho v_x Y_{n_{\textrm{sp}}}  \end{bmatrix} \vec{i} +  \begin{bmatrix} \rho v_y \\ \rho v_x v_y \\ \rho v_y^2 + p  \\ \rho v_y e+ p v_y \\ \rho v_y Y_1 \\ \vdots \\ \rho v_y Y_{n_{\textrm{sp}}} \end{bmatrix} \vec{j},
\qquad
\vec{K}_v =
\begin{bmatrix}  0 \\ \tau_{xx} \\ \tau_{yx} \\ \tau_{xx}v_x +  \tau_{yx}v_y - j_x^q \\ -j_{1,x}^m\\ \vdots \\ -j_{n_{\textrm{sp}},x}^m\end{bmatrix} \vec{i} +
\begin{bmatrix}  0 \\ \tau_{xy} \\ \tau_{yy} \\  \tau_{xy}v_x +  \tau_{yy}v_y  - j_y^q \\ -j_{1,y}^m\\ \vdots \\ -j_{n_{\textrm{sp}},y}^m\end{bmatrix} \vec{j}.
\end{equation*}
The two-dimensional viscous shear tensor is defined as
\begin{equation*} 
\tau
= \begin{bmatrix} \tau_{xx} & \tau_{xy} \\ \tau_{xy} & \tau_{yy} \end{bmatrix}
= \hat{\mu} \begin{bmatrix} \frac{1}{3} \frac{\partial v_x}{\partial x} & \frac{\partial v_x}{\partial y}  + \frac{\partial v_y}{\partial x} \\ \frac{\partial v_y}{\partial x} + \frac{\partial v_x}{\partial y}& \frac{1}{3} \frac{\partial v_y}{\partial y} \end{bmatrix},
\end{equation*}
where $\hat{\mu}$ is the mixture viscosity coefficient.
The diffusive heat flux vector is defined as
\begin{equation}
\vec{j}^q = \left [ j_x^q \quad j_y^q \right ]^{\top} = -\kappa \nabla T + \rho \sum_{l = 1}^{n_{\textrm{sp}}} D_{l} h_l \nabla Y_l ,
\label{eq:heatflux}
\end{equation}
where $\kappa$ quantifies thermal conductivity and $D_l$ is the diffusion coefficient for the $l$th species into the mixture, which is an approximation used to model the multi-component diffusion as the binary diffusion of each species into a mixture. The two terms in the definition of the heat flux (Eq.~\eqref{eq:heatflux}) represent heat transfer due to conductivity and species diffusion. 
The diffusive mass flux vector of species $l$ is modeled as
\begin{equation*}
\vec{j}^m_l = \begin{bmatrix}j_{l,x}^m \quad j_{l,y}^m\end{bmatrix}^{\top} = \begin{bmatrix}\rho D_l \frac{\partial Y_l}{\partial x} \quad \rho D_l \frac{\partial Y_l}{\partial y }\end{bmatrix}^{\top}.
\end{equation*}
The source term, $\vec{S}$ in Eq.~\eqref{eq:ns_conservative} is
\begin{equation}
\vec{S}~=~ \left [ 0\quad 0 \quad 0 \quad 0 \quad \dot{\omega}_1 \quad \dots \quad \dot{ \omega}_{n_\textrm{sp}} \right ]^{\top} \label{eq:defS}
\end{equation}
and is defined by considering a 1-step combustion reaction governed by
$$ \textrm{CH}_4 + 2\textrm{O}_2 \rightarrow \textrm{CO}_2 + 2\textrm{H}_2\textrm{O},$$
as presented in \cite{combustion_chemistry}, with $n_{\textrm{sp}} = 4$. The corresponding general stoichiometric equation is defined as $ 0 = \sum_{l=1}^{n_{\textrm{sp}}} \nu_l \chi_l,$
where $\chi_1 = \textrm{CH}_4, \chi_2 = \textrm{O}_2, \chi_3 = \textrm{CO}_2$, $\chi_4 = $ H$_2$O and $\nu_l$ is the net stoichiometric coefficients of each species with $\nu_1 = -1, \nu_2 = -2, \nu_3 = 1$ and $\nu_4 = 2$.
The molar concentration of the $l$th species is denoted by $c_l$. In our case, $l \in \{1,2,3,4\}$, so $c_1 = [\textrm{CH}_4], \ c_2 = [\textrm{O}_2], \ c_3= [\textrm{CO}_2],$ and $ c_4 = [\textrm{H}_2\textrm{O}]$ are the molar concentrations. Here, we use the standard bracket notation $[\cdot]$ to indicate molar concentration of a species.
The general relationship between a species molar concentration, $c_l$, and a species mass fraction, $Y_l$, is
\begin{equation}
Y_l = \frac{c_l M_l}{\rho}, \label{eq:moletomass}
\end{equation}
where $Y_1$ is the mass fraction of CH$_4$, $Y_2$ is the mass fraction of O$_2$, $Y_3$ is the mass fraction of CO$_2$ and $Y_4$ is the mass fraction of H$_2$O.
The production rate of the $l$th species in the source term $\vec{S}$  in Eq.~\ref{eq:defS} is modeled as
\begin{equation}
\dot {\omega}_l = M_l \frac{\partial c_l^{\text{reaction}}}{\partial t} = \nu_l \Gamma_r,
 \label{eq:omega_dot}
\end{equation}
where $c_l^{\text{reaction}}$ are chemical reaction source terms whose dynamics are described below and $\Gamma_r$ is the reaction rate. The molar mass of CH$_4$ is $M_1 = 16.04 \frac{\textrm{g}}{\textrm{mol}}$, the molar mass of O$_2$ is $M_2 = 32.0 \frac{\textrm{g}}{\textrm{mol}}$, the molar mass of CO$_2$ is $M_3 = 44.01 \frac{\textrm{g}}{\textrm{mol}}$, and the molar mass of H$_2$O is $M_4 =18.0 \frac{\textrm{g}}{\textrm{mol}}$.

%
%
The reaction rate is approximated by
\begin{equation*}
\Gamma_r = k      \prod_{l=1}^{n_{\textrm{reactant}}} c_l^{o_l},
\end{equation*}
where $n_{\textrm{reactant}} = 2$ is the number of reactants, $k$ is the rate coefficient and $o_l$ is the reaction order of the $l$th reactant. In our case $o_1 = 0.2$ and $o_2 = 1.3$. The rate coefficient, $k$, is described by the Arrhenius equation as
\begin{equation} \label{eq:Arrhenius_Coeff}
k =  A\exp\left(\frac{-E_a}{R_{\textrm{u}} T}\right),
\end{equation}
where $R_{\textrm{u}}  = 8.314 \frac{\text{J}}{\text{mol K}}$ is the universal gas constant, $A = 2\times 10^{10}$ is the pre-exponential constant and $E_a = 2.025\times 10^5$ is the energy required to reach a chemical reaction, measured in Joules and referred to as the activation energy.
In this work, we use the ideal gas state equation that relates density and pressure to temperature
\begin{equation}
\rho = \frac{p}{R T}, \label{eq:idealGas}
\end{equation}
where $R = \frac{R_u}{M}$ and $M=\left ( \sum^{n_{sp}}_{l=1}\left(\frac{Y_l}{M_l}\right)\right ) ^{-1}$ is the average molar mass of the mixtures. {\Rev Thus, we can obtain temperature via $T = \frac{p}{\rho \ R(Y_l)}$ from the states  $\rho,p, Y_l$.}

At the downstream end of the combustor, we impose a non-reflecting boundary condition while maintaining the chamber pressure via
\begin{equation}
p_{\textrm{back}}(t) = p_{\textrm{back,ref}} [1+A\sin(2\pi ft)],
\label{eq:pback}
\end{equation} 
where $p_{\textrm{back,ref}} = 1.0\times 10^6$ Pa, $A = 0.1$ and $f = 5000$Hz.
The top and bottom wall boundary conditions are no-slip conditions, and for the upstream boundary we impose constant mass flow at the inlets.

\subsection{Numerical model} \label{sec:GEMS_discretization}
GEMS uses the finite volume method to discretize the {\Rev conservation equations~\eqref{eq:ns_conservative}. The primitive variables $\vec{q}_p=[p \ v_x  \ v_y \ T \ Y_{1} \ldots Y_{n_{\textrm{sp}}}]^{\top}$ are chosen as solution variables in GEMS, since they allow for easier computation of thermal properties and provide more flexibility when extending to complex fluid problems like liquid and supercritical fluids}. 
For a spatial discretization with $n_x$ cells, this results in a $dn_x$-dimensional system of nonlinear ordinary differential equations (ODEs)
\begin{equation}
\frac{\text{d}\bq}{\text{d}t} = \mathbf{G}(\bq, \bu(t)), \qquad \bq(0) = \bq_0, \label{eq:F}
\end{equation}
for $0<t\leq T$, where $d$ is the number of unknowns in the PDE governing equations and here $d=8$ (four flow variables and four species concentrations). In Eq.~(\ref{eq:F}), $\bq(t) \in \mathbb{R}^{dn_x}$ is the discretized state vector at time $t$ (for GEMS, it is the discretization of the primitive variables $\vec{q}_p=[p \ v_x  \ v_y \ T \ Y_{1} \ldots Y_{n_{\textrm{sp}}}]^{\top}$), $\bq_0$ are the specified initial conditions, and $\frac{\text{d}\bq}{\text{d}t}$ is the time derivative of the state vector at time $t$.  The $m$ inputs $\bu(t) \in \mathbb{R}^m$ arise from the time-dependent boundary condition, defined in Eq.~(\ref{eq:pback}), applied at the combustor downstream end. The nonlinear function $\mathbf{G}: \mathbb{R}^{dn_x} \times \mathbb{R}^m \rightarrow \mathbb{R}^{dn_x}$ maps the discretized states $\bq$ and the input $\bu$ to the state time derivatives, representing the spatial discretization of the governing equations described in Section~\ref{sec:governing}.

Solving these high-dimensional nonlinear ODEs is expensive, motivating the derivation of a ROM that can yield approximate solutions at reduced cost. The nonlinear multi-scale dynamics represented by these equations makes this a challenging task. 
To maintain computational efficiency in the ROM, state-of-the-art nonlinear model reduction methods combine POD with a sparse interpolation method (often called hyperreduction)  by evaluating the nonlinear functions only at a select number of points. For instance, POD together with the discrete empirical interpolation method (DEIM) has had some success, but also encountered problems in combustion applications~\cite{huang2018exploration} . 
Of particular challenge is the need to include a large number of interpolation points in the POD-DEIM approximation, which means that the ROM loses its computational efficiency. Robustness and stability of the POD-DEIM models is also a challenge~\cite{huang2018challenges}. In the next section, we present a different approach that uses non-intrusive ROM learning to enable variable transformations that expose system structure. This structure is then exploited in the derivation of the ROM and removes the need for the DEIM approximation.

\section{Non-intrusive learning of a combustion reduced model} \label{sec:rom}
This section presents our approach to learn ROMs for the unsteady combustion dynamics simulation from GEMS. Section~\ref{sec:poly_FOM} writes a general nonlinear system in a form that exposes the underlying structure of the governing equations and shows how projection preserves that structure. Section~\ref{sec:opinf} presents the operator inference approach from~\cite{Peherstorfer16DataDriven}, which learns structured ROM operators from simulation data. Section~\ref{sec:transform} describes variable transformations that lead to the desired polynomial structure for the combustion governing equations presented in Section~\ref{sec:combModel}. These transformations yield the structure needed to apply the operator inference approach.

\subsection{Projection preserves polynomial structure in the governing equations} \label{sec:poly_FOM}

Consider a large-scale sytem of nonlinear ODEs written in polynomial form
\begin{equation}
\frac{\text{d}\bq}{\text{d}t} = {\bA} \bq + {\bH} (\bq \otimes \bq) + {\mathbf{C}} (\bq\otimes \bq \otimes \bq) + {\bB}\bu + {\mathbf{c}} + \textrm{HOT}. \label{eq:poly_FOM}
\end{equation}
Relating this equation to the general nonlinear system in Eq.~(\ref{eq:F}), we see that ${\bA} {\bq}$ are the terms in $\mathbf{G(\cdot)}$ that are linear in the state $\bq$, with ${\bA}\in \mathbb{R}^{dn_x \times dn_x}$; ${\bH} ({\bq}\otimes {\bq})$ are the terms in $\mathbf{G(\cdot)}$ that are quadratic in $\bq$, with ${\bH} \in \mathbb{R}^{dn_x \times (dn_x)^2}$; ${\mathbf{C}} ({\bq}\otimes {\bq}\otimes {\bq})$ are the terms in $\mathbf{G(\cdot)}$ that are cubic in $\bq$, with ${\mathbf{C}} \in \mathbb{R}^{dn_x \times (dn_x)^3}$; ${\bB}\bu$ are the terms in $\mathbf{G(\cdot)}$ that are linear in the input $\bu$, with ${\bB}\in \mathbb{R}^{dn_x \times m}$; and $\mathbf{c}\in \mathbb{R}^{dn_x}$ are constant terms in $\mathbf{G(\cdot)}$ that do not depend on state or input. The abbreviation ``HOT'' in Eq.~\eqref{eq:poly_FOM} denotes higher-order terms, and represents terms that are quartic and higher order, as well as any other nonlinear terms that cannot be represented in polynomial form.

We emphasize that we are not (yet) introducing approximations---rather, we are explicitly writing out the discretized equations in the form \eqref{eq:poly_FOM} to expose the system structure that arises from the form of the terms in the governing PDEs. For example, a term such as $\frac{\partial}{\partial x} \rho v_x$ in Eq.~(\ref{eq:ns_conservative}) is linear in the state $\rho v_x$, while a term such as $\frac{\partial}{\partial x} \rho v_x Y_1$ is quadratic in the states $\rho v_x$ and $Y_1$. Also note that the term $\frac{\partial}{\partial x} \rho v_x$ is quadratic in the states $\rho$ and  $v_x$, highlighting the important point that the structure of the nonlinear model depends on the particular choice of state variables.

A projection-based ROM of Eq.~\eqref{eq:poly_FOM} preserves the polynomial structure. Approximating the high-dimensional state $\bq$ in a low-dimensional basis $\bV\in \mathbb{R}^{dn_x \times r}$, with $r\ll dn_x$, we write $\bq \approx \bV\widehat{\bq}$. Using a Galerkin projection, this yields the ROM of Eq.~\eqref{eq:poly_FOM} as
\begin{equation}
\frac{\text{d}\widehat{\bq}}{\text{d}t} = \widehat{\bA} \widehat{\bq}+ \widehat{\bH} (\widehat{\bq}\otimes \widehat{\bq}) + \widehat{\mathbf{C}} (\widehat{\bq}\otimes \widehat{\bq}\otimes \widehat{\bq})+ \widehat{\bB}\bu+ \widehat{\mathbf{c}} + \textrm{HOT},
\label{eq:poly_ROM}
\end{equation}
where $\widehat{\bA}=\bV^\top \bA \bV \in \mathbb{R}^{r \times r} $,  $\widehat{\bH} = \bV^\top \bH (\bV \otimes \bV) \in \mathbb{R}^{r \times r^2}$, $\widehat{\mathbf{C}} = \bV^\top \mathbf{C} (\bV \otimes \bV \otimes \bV) \in \mathbb{R}^{r \times r^3}$, and $\widehat{\bB} = \bV^\top \bB \in \mathbb{R}^{r\times m}$ are the ROM operators corresponding respectively to $\bA$, $\bH$, $\mathbf{C}$, and $\bB$, and $\widehat{\mathbf{c}} = \bV^\top \mathbf{c} \in \mathbb{R}^r$ is a constant vector. We note again that projection preserves polynomial structure, that is, \eqref{eq:poly_ROM} has the same polynomial form as \eqref{eq:poly_FOM}, but in the reduced subspace defined by $\bV$. 

In what follows, we will work with a quadratic system in order to simplify notation. We note that the least squares learning approach described below applies directly to cubic, quartic and all higher-order polynomial terms (although it should be noted that the number of elements in the ROM operators scales with $r^4$ for the cubic operator, $r^5$ for the quartic operator, etc.). 
{\Rev Higher-order terms often exhibit significant block-sparsity that can be exploited in numerical implementations, which limits the growth of computational cost to solve the ROM.}
For terms in the governing equations that are not in polynomial form (such as terms involving $\frac{1}{\rho}$, and the Arrhenius reaction terms) we discuss in Section~\ref{sec:transform} the introduction of variable transformations and auxiliary variables via the process of lifting \cite{gu2011,KW18nonlinearMORliftingPOD} to convert these terms to polynomial form.

\subsection{Operator inference for learning reduced models} \label{sec:opinf}
Here we summarize the steps of the operator inference approach from~\cite{Peherstorfer16DataDriven}.
First, we collect $K$ snapshots of the state  by solving the high-fidelity model. We store the snapshots and the inputs used to generate them in the matrices:
\begin{equation*}
\bQ = [\bq_0 \ \dots \bq_K] \in \mathbb{R}^{dn_x \times K}, \qquad \mathbf{U}= [\bu_0,\ \dots, \ \bu_K] \in \mathbb{R}^{m \times K},
\end{equation*}
where $\bu_i \equiv \bu(t_i)$ and $\bq_i \equiv \bq(t_i)$ with $0 = t_0 < t_1< \dots < t_K = T$. In general, $dn_x \gg K$, so the matrix $\bQ$ is tall and skinny.
Second, we identify the low-dimensional subspace in which we will learn the ROM. In this work, we use the POD to define the low-dimensional subspace, by computing the singular value decomposition of the snapshot matrix
$$
\bQ = \bV\boldsymbol{\Sigma}\mathbf{W}^{\top},
$$
where $\bV\in \mathbb{R}^{dn_x \times K}$, $\boldsymbol{\Sigma} \in \mathbb{R}^{K \times K}$ and $\mathbf{W}  \in \mathbb{R}^{K \times K}$. The $r \ll dn_x$  dimensional POD basis, $\bV_r = [\bv_1,...,\bv_r]$, is given by the first $r$ columns of $\bV$.
Third, we project the state snapshot data onto the POD subspace spanned by the columns of $\bV_r$ and obtain the reduced snapshot matrices
\begin{equation*}
\widehat{\bQ} = \bV^{\top}_r \bQ = [\widehat{\bq}_0 \quad \dots \quad \widehat{\bq}_K] \in \mathbb{R}^{r \times K},
\qquad
\dot{\widehat{\bQ}} = [ \dot{\widehat{\bq}}_0\quad \dot{\widehat{\bq}}_1 \quad \dots \quad  \dot{\widehat{\bq}}_K]\in \mathbb{R}^{r \times K},
\end{equation*}
where the columns of $\dot{\widehat{\bQ}}$ are computed from $\widehat{\bQ}$ using any time derivative approximation (see, e.g.,~\cite{martins2013review,knowles2014methodsDifferentiation,chartrand2017numericalDifferentiation}), or can be obtained---if available---by collecting and projecting snapshots of $\mathbf{G}(\bq_i, \bu_i)$.

Operator inference solves a least squares problem to find the reduced operators that yield the ROM that best matches the projected snapshot data in a minimum residual sense. For the quadratic ROM
\begin{equation}
\dot{\widehat{\bq}} = \widehat{\bA} \widehat{\bq}+ \widehat{\bH} (\widehat{\bq}\otimes \widehat{\bq}) + \widehat{\bB}\bu+ \widehat{\mathbf{c}},
\label{eq:quadratic_system_reduced}
\end{equation}
operator inference solves the least squares problem
\begin{equation*}
\min_{ \widehat{\bA} \in \mathbb{R}^{r \times r} , \widehat{\bH} \in \mathbb{R}^{r \times r^2}, \widehat{\bB}  \in \mathbb{R}^{r \times m}, \bhc \in \mathbb{R}^r}
\left \Vert   \widehat{\bQ}^{\top}\widehat{\bA}^{\top} + (\widehat{\bQ} \otimes \widehat{\bQ})^{\top} \widehat{\bH}^{\top}  + \mathbf{U}^{\top}\widehat{\bB}^{\top} + \mathbf{1}_K \bhc^{\top}- \dot{\widehat{\bQ}}^{\top} \right \Vert^2_2,
\end{equation*}
where $\mathbf{1}_K \in \mathbb{R}^{K}$ is the length $K$ column vector with all entries set to unity.
Note that this least squares problem is linear in the coefficients of the unknown ROM operators
$\widehat{\bA}$, $\widehat{\bH}$, $\widehat{\bB}$ and $\widehat{\mathbf{c}}$. Also note that the operator inference approach permits us to compute the ROM operators  $\widehat{\bA}$, $\widehat{\bH}$, $\widehat{\bB}$ and $\widehat{\mathbf{c}}$ without needing explicit access to the original high-dimensional operators ${\bA}$, ${\bH}$, ${\bB}$ and ${\mathbf{c}}$.

We combine the unknown operators of Eq.~\eqref{eq:quadratic_system_reduced} in the matrix
\begin{equation*}
\mathbf{O} = [\widehat{\bA}\quad \widehat{\bH}\quad \widehat{\bB}\quad \bhc] \in \mathbb{R}^{r \times(r+r^2+m + 1)},
\end{equation*}
and the known low-dimensional data in the data matrix
\begin{equation} \label{eq:datamatrix}
\mathbf{D} = \begin{bmatrix} \widehat{\bQ}^{\top} \quad (\widehat{\bQ} \otimes \widehat{\bQ})^{\top} \quad \mathbf{U}^{\top}\quad \mathbf{1}_K\end{bmatrix} \in \mathbb{R}^{K \times (r + r^2+ m + 1)},
\end{equation}
and then solve the minimization problem
\begin{align}
\min_{\mathbf{O}\in \mathbb{R}^{r \times (r+r^2+m + 1)} } \left \Vert \mathbf{D} \mathbf{O}^{\top} - \dot{\widehat{\bQ}}^{\top} \right \Vert^2_2.
\label{eq:min_OD}
\end{align}
{\Rev For $K>r+r^2+m+1$ this overdetermined linear least-squares problem has a unique solution~\cite[Sec. 5.3]{golub96matrix}.}
It was proven in \cite{Peherstorfer16DataDriven} that Eq.~\eqref{eq:min_OD} can be written as $r$ independent least squares problems of the form $ \min_{\mathbf{o}_i \in \mathbb{R}^{r+r^2+m+1}} \left \Vert\mathbf{D} \mathbf{o}_i - \mathbf{r}_i \right \Vert_2^2$, for $i = 1,\dots,r$, where $\mathbf{o}_i$ is a column of $\mathbf{O}^{\top}$ (row of $\mathbf{O}$) and $\mathbf{r}_i$ is a column of $\dot{\widehat{\bQ}}^{\top}$. This makes the operator inference approach efficient and scalable.

Regularization becomes necessary to avoid overfitting and to infer operators that produce a stable ROM. In this work, we use an L$_2$ regularization penalty on the off-diagonal elements of the operator $\widehat{\bA}$ and on all elements of the remaining operators. With this regularization, our least squares problem becomes
\begin{equation}
\min_{\mathbf{o}_i \in \mathbf{R}^{r + r^2 + m+ 1}} \left \Vert \mathbf{D} \mathbf{o}_i - \mathbf{r}_i \right \Vert_2^2 + \lambda \left \Vert \mathbf{P}_i \mathbf{o}_i \right \Vert_2^2 \quad \textrm{for } i=1,\ldots,r,
\label{eq:final_minimization}
\end{equation}
where $\lambda$ is the regularization parameter and $\mathbf{P}_i$ is the $r+r^2+m+1$ identity matrix with the $i$th diagonal set to zero so that we avoid regularizing the diagonal elements of $\bA$. It should be noted that the regularization parameter, $\lambda$, is problem specific and should be chosen accordingly. In Section~\ref{sec:implementation}, we discuss details of the operator inference implementation, a method for selecting $\lambda$, and the removal of redundant terms in the least squares problem in Eq.~\eqref{eq:final_minimization}.

\subsection{A structure-exploiting ROM learning formulation for GEMS} \label{sec:transform}

A key contribution of this work is to recognize that the non-intrusive operator inference approach gives us complete flexibility in the set of physical variables we work with to define the ROM. We can identify choices of physical variables that expose the desired polynomial structure in the governing equations, and then extract snapshots for those variables by applying transformations to the snapshot data---we do not need to make any modifications to the high-fidelity CFD simulation model itself. In theory, a classical intrusive ROM approach could work with transformed variables (e.g., in the work of \cite{balajewicz2016minimal}); however, this would involve rewriting the high-fidelity simulator, a task that would be not only time-consuming but also fraught with mathematical pitfalls, especially for unusual choices of variables. This is where the data-driven perspective of machine learning becomes extremely valuable.

The Euler equations admit a quadratic representation in the specific volume variables; in that case, a transformation of the snapshots from conservative (or primitive) variables to specific volume variables can be exploited to create quadratic ROMs \cite{elizabeth}. Other PDEs may not admit polynomial structure via such straightforward transformations, but the process of \emph{lifting} the equations via the introduction of new auxiliary variables can produce a set of coordinates in which the governing equations become polynomial in the lifted state \cite{gu2011,KW18nonlinearMORliftingPOD,KW_BT_liftedSystems_2019}. For example, the tubular reactor example of \cite{KW18nonlinearMORliftingPOD} includes Arrhenius-type reaction terms similar to those in Eq.~\eqref{eq:Arrhenius_Coeff}. The introduction of auxiliary variables permits the governing equations to be written equivalently with quartic nonlinearity in the lifted variables.\footnote{The Arrhenius reaction terms can be lifted further to quadratic form, but then require the inclusion of algebraic constraints, which makes the model reduction task more difficult, see \cite{KW18nonlinearMORliftingPOD}.}

Lifting to polynomial form for the GEMS equations described in Section~\ref{sec:combModel} is made difficult by several of the terms, in particular through some of the gas thermal properties such as the nonlinear dependence of enthalpy on temperature. A complete lifting that converts all equations to a polynomial form is possible, but would require the introduction of a large number of auxiliary variables and would also result in the introduction of some algebraic equations. However, as the analysis below shows, the GEMS governing equations admit a transformation for which many terms in the governing equations take polynomial form when we use the variables
\begin{eqnarray}
\vec{q}_{L} & = \begin{bmatrix} p & v_x  & v_y & \xi & c_{1}& c_2 & c_3 & c_4 \end{bmatrix}^{\top}.
\label{eq:transform_var}
\end{eqnarray}
Here $\xi = \frac{1}{\rho}$ is the specific volume, and recall that $c_1 = [\textrm{CH}_4], \ c_2 = [\textrm{O}_2], \ c_3= [\textrm{CO}_2],$ and $ c_4 = [\textrm{H}_2\textrm{O}]$ are the molar concentrations with $c_l = \frac{\rho Y_l}{M_l}$. 

Below, we derive the governing PDEs for specific volume $\xi$ and velocities $v_x, v_y$. These three governing PDEs all turn out to be quadratic in the learning variables $\vec{q}_L$.
In Appendix~A we present the lifting transformations for the source term dynamics $c_l^\text{reaction}$ in the vector $\vec{S}$ in Eq.~\eqref{eq:defS}. In Appendix~B we derive the equations governing the pressure $p$ and the species molar concentrations $c_i$. These equations have some terms that are not polynomial in the chosen learning variables $\vec{q}_L$.

To keep notation clean in application of the chain rule, let the conservative variables be denoted as $g_1 = \rho, \ g_2 = \rho v_x, \ g_3 = \rho v_y, \ g_4 = \rho e, \ g_5 = \rho Y_1, \ g_6 = \rho Y_2, \ g_7 = \rho Y_3, \ g_8 = \rho Y_4$.
Throughout, we frequently use the relationship
\begin{equation} \label{eq:drhodx}
\frac{\partial \xi }{\partial x}  = \frac{\partial }{\partial x} \frac{1}{\rho} = -\frac{1}{\rho^2} \frac{\partial \rho }{\partial x} = -\xi^2 \frac{\partial \rho}{\partial x},
\end{equation}
and similarly for $\frac{\partial \xi }{\partial y}$. Note also that we are assuming the existence of these partial derivatives, that is, we do not consider the case of problems with discontinuities.\\

\noindent \textbf{Specific Volume $\xi = 1/\rho$}. We use the constitutive relationship for the density $\rho$ in Eq.~\eqref{eq:ns_conservative} in the derivation:
\begin{equation*}
\frac{\partial \xi}{\partial t}
=  \frac{\partial}{\partial t} \frac{1}{\rho}
= - \frac{1}{\rho^2} \dot{\rho}
=  \xi^2 \nabla \cdot \left ( \rho v_x\vec{i} + \rho v_y \vec{j} \right )
= \xi^2\left [ \frac{\partial \rho}{\partial x} v_x + \rho \frac{\partial  v_x}{\partial x} \right ] + \xi^2\left [v_y \frac{\partial \rho}{\partial y} + \rho \frac{\partial v_y}{\partial y}  \right ].
\end{equation*}
Inserting Eq.~\eqref{eq:drhodx} into the above, we obtain
\begin{equation*}
\frac{\partial \xi}{\partial t}
=  - \frac{\partial \xi}{\partial x} v_x + \xi\frac{\partial  v_x}{\partial x}  - v_y \frac{\partial \xi}{\partial y} + \xi \frac{\partial v_y}{\partial y},
\end{equation*}
which is \underline{quadratic} in the learning variables $\xi, v_x, v_y$. \\

\noindent \textbf{Velocities $v_x,v_y$:} We have
\begin{equation*}
\frac{\partial v_x }{\partial t}   = \frac{\partial}{\partial t} \frac{g_2}{g_1} = \frac{1}{g_1} \dot{g}_2 - g_2 \frac{1}{g_1^2} \dot{g}_1 = \frac{1}{\rho} \dot{g}_2 - \frac{v_x}{\rho} \dot{g}_1,
\end{equation*}
and from Eq.~\eqref{eq:ns_conservative} we have that  $\dot{g}_1 = \dot{\rho} = - \nabla \cdot \left ( \rho v_x\vec{i} + \rho v_y \vec{j} \right ) $ as well as $\dot{g}_2 = \dot{(\rho v_x)} =\nabla \cdot \left ( - (\rho v_x^2 + p)\vec{i} - (\rho v_x v_y) \vec{j} + \tau_{xx} \vec{i} + \tau_{xy}\vec{j} \right ) $. Thus, we obtain
\begin{align*}
\frac{\partial v_x}{\partial t}
& = \frac{1}{\rho} \nabla \cdot \left ( - (\rho v_x^2 + p)\vec{i} - (\rho v_x v_y) \vec{j} + \tau_{xx} \vec{i} + \tau_{xy}\vec{j} \right ) + \frac{v_x}{\rho} \nabla \cdot \left ( \rho v_x\vec{i} + \rho v_y \vec{j} \right ), \\
& = -\xi \frac{\partial \rho}{\partial x} v_x^2 -  \frac{\partial v_x^2}{\partial x} - \xi \frac{\partial p}{\partial x} - \xi \frac{\partial \rho}{\partial y} v_x v_y - \frac{\partial v_x v_y}{\partial y} + \xi \left (\frac{\partial \tau_{xx}}{\partial x}  + \frac{\partial \tau_{xy}}{\partial y} \right ) + v_x^2 \xi \frac{\partial \rho}{\partial x} + v_x \frac{\partial v_x}{\partial x} + \xi v_x v_y \frac{\partial \rho}{\partial y} + v_x \frac{\partial v_y}{\partial y}\\
& = -  \frac{\partial v_x^2}{\partial x} - \xi \frac{\partial p}{\partial x} - \frac{\partial v_x v_y}{\partial y} + \xi \left (\frac{\partial \tau_{xx}}{\partial x}  + \frac{\partial \tau_{xy}}{\partial y} \right ) + v_x \frac{\partial v_x}{\partial x}  + v_x \frac{\partial v_y}{\partial y}\\
& = - \xi \frac{\partial p}{\partial x} - v_y\frac{\partial v_x}{\partial y} + \xi \left (\frac{\partial \tau_{xx}}{\partial x}  + \frac{\partial \tau_{xy}}{\partial y} \right ) - v_x \frac{\partial v_x}{\partial x}
\end{align*}
and we get a similar expression for $\frac{\partial v_y}{\partial t}$. Both dynamics are
\underline{quadratic} in the learning variables $p,v_x, v_y, \xi$.\\

As noted above, Appendix~A  and Appendix~B present the derivations for the chemical source terms, pressure, and chemical species.

\section{Numerical Results} \label{sec:numerics}
We now apply the variable transformations and operator inference framework to learn a predictive ROM from GEMS high-fidelity combustion simulation data.\footnote{Code for the operator inference framework is available at \url{https://test.pypi.org/project/operator-inference/} in Python and \url{https://github.com/elizqian/operator-inference} in Matlab.} Section~\ref{sec:GEMSdata_results} describes the problem setup and GEMS dataset. Section~\ref{sec:implementation} discusses implementation details and Section~\ref{sec:combustion_results}  presents our numerical results. Additional numerical results can be found in~\cite{Swischuk2019thesis}.

\subsection{GEMS Dataset} \label{sec:GEMSdata_results}
The computational domain shown in Figure~\ref{fig:combustor2} is discretized with $n_x = 38523$ spatial discretization points. Each CFD state solution thus has dimension $dn_x = 308184$.
The problem considered here has fuel and oxidizer input streams with constant mass flow rates of $5.0 \frac{\text{kg}}{\text{s}}$ and $0.37 \frac{\text{kg}}{\text{s}}$, respectively. The fuel is composed of gaseous methane and the oxidizer is 42\% gaseous O$_2$ and 58\% gaseous H$_2$O, as described in \cite{cheng}. The forcing input Eq.~\eqref{eq:pback} is applied at the right side of the domain.
{\Rev For this simulation, the resulting Reynolds number is about 10,000, defined as $Re=\frac{\rho{v_x L}}{\mu}$ where the density $\rho$, horizontal velocity $v_x$ and viscosity $\mu$ are evaluated at the inlet of the oxidizer post ($x=-0.04 $m in Fig.~\ref{fig:combustor2}), and the characteristic length $L$ is defined as the height of the oxidizer inlet. The highest Mach number is $\approx0.25$ and is evaluated inside the oxidizer post (from $x=-0.04$m to $0$m in Fig.~\ref{fig:combustor2}).
}

To generate training data, GEMS is simulated for a time duration of 1ms with a time step size of  $\Delta t = 1\times 10^{-7}$s  resulting in $K =10000$ snapshots.
The GEMS output is transformed to the variables given in Eq.~\eqref{eq:transform_var}.
The recorded snapshot matrix is thus
\begin{equation*}
\bQ = [\bq_0\quad \bq_1 \quad \dots \quad \bq_K] \in \mathbb{R}^{dn_x \times K} = \mathbb{R}^{308184 \times 10000}
\end{equation*}
Our numerical experiments were parallelized on a  cluster with two computing nodes. Each node has two 10-core Intel Xeon-E5 processors (20 cores per node) and 128 GB RAM. The training data generation took approximately 200h in CPU time for the 1ms, 10000 snapshots of high-fidelity CFD data.

The range of variable values for the training data is shown in Table~\ref{table:range_of_values1}. Note that the data covers a wide range of scales. Pressure is of the order $10^6$ while species concentrations can be as low as 10$^{-12}$. This large scaling difference presents a challenge when learning models from data.
%
To deal with the numerical issues related to large differences in  scaling and small species concentrations and velocities, we scale each variable to the interval $[-1,1]$. Variables are scaled before computing the POD basis and projecting the data.
%
\begin{table}[h!]
\begin{center}
\caption{Range of variable values for GEMS data.}
	\label{table:range_of_values1}
		\begin{tabular}{l | r | r | r }
			State variable & Minimum & Mean & Maximum \\ \hline
			Pressure $p$ in Pa & $9.226 \times 10^5$ & $1.142 \times 10^6 $ & $1.433\times 10^6$  \\
			Velocity $v_x$ in $\frac{\text{m}}{\text{s}}$ & -222.930 & 69.637 & 307.147  \\
			Velocity $v_y$ in $\frac{\text{m}}{\text{s}}$ & -206.990 & 1.304 & 186.548  \\
			Specific volume $\xi={\rho}^{-1}$ in $\frac{\text{m}^3}{\text{kg}}$ & $0.106$ & $0.333$ & $1.021$ \\
			Molar concentration $[{CH}_4]$  & $0.000$ & $0.035$ & $0.586$  \\
			Molar concentration $[{O}_2]$   & $0.000$ & $0.038$ & $0.066$ \\
			Molar concentration $[{CO}_2]$ & $0.000$ & $0.002$ & $0.008$ \\
			Molar concentration $[{H}_2 O]$ & $0.000$ & $0.104$ & $0.161$
			%
		\end{tabular}%
\end{center}		
\end{table}

To obtain snapshots of the projected state time derivative, we approximate the derivative with a five-point approximation $\dot{\widehat{\bq}}_i = (-{\widehat{\bq}}_{i+2} + 8 {\widehat{\bq}}_{i+1} - 8{\widehat{\bq}}_{i-1} +{\widehat{\bq}}_{i-2} )/(12\Delta t)$. This approximation is fourth order accurate. The first two and last two time derivatives are computed using first-order forward and backward Euler approximations, respectively.

\subsection{Learning a quadratic reduced-order model} \label{sec:implementation}
To learn the operators of the quadratic ROM, we solve the regularized least squares problem shown in Eq.~\eqref{eq:final_minimization}. {\Rev  We are using numpy's least squares solve \texttt{numpy.linalg.lstsq}. The algorithm is based on the LAPACK routine xGELSD. That routine is based on the SVD, which typically provides a stable implementation. }
In what follows next, we describe several important implementation details.

\subsubsection{Singular value decomposition implementation and POD basis selection}
Due to the large size of this dataset, we implement the randomized singular value decomposition algorithm, introduced in \cite{randomizedSVD}, to compute the leading 500 singular values and vectors of the snapshot matrix. The randomized singular value decomposition algorithm can be implemented in a scalable way for large datasets {\Rev as the data does not have to be read into single memory all at once}.
The POD basis is chosen as the $r$ leading left singular vectors. The dimension $r$ is typically chosen so that the cumulative energy contained in the subspace is greater than a user specified tolerance $\epsilon$, i.e.,
$$
\frac{{ \sum_{k=1}^r \sigma_k^2}}{{ \sum_{k=1}^{dn_x} \sigma_k^2}} > \epsilon,
$$
where $\sigma_k^2$ are the squared singular values of the data matrix $\bQ$. To guide the choice of $r$ we also use the relative projection error
\begin{equation}
\mathcal{E}_{\textrm{proj}} = \frac{\Vert \bQ - \bV_r \bV_r^{\top}\bQ  \Vert_F^2}{\Vert \bQ \Vert^2_F} = 1 - \frac{{ \sum_{k=1}^r \sigma_k^2}}{ \sum_{k=1}^{dn_x} \sigma_k^2}
\label{eq:projection_error}
\end{equation}

\subsubsection{Removing redundant terms in least squares problem}
{ \Rev
There are redundant terms that arise in the Kronecker product $\widehat{\bQ} \otimes \widehat{\bQ}$ in Eq.~\eqref{eq:datamatrix}, which can cause the least squares problem to become ill-posed. To see this, consider $q \otimes q$ with $q=[q_1 \ q_2]^T$, i.e., we have $q \otimes q = [q_1^2 \ \ q_2q_1 \ \ q_1q_2 \ \  q_2^2]$. Since we know where the repeated terms occur in the product we merely need to remove the redundant (repeated) terms before we solve the least-squares problem. 
 }
Thus, the Kronecker product is replaced with the term
$$
\widehat{\bQ}^2 = \left  [\widehat{\bq}^2_0 \quad \widehat{\bq}^2_1 \quad \dots\quad \widehat{\bq}^2_K \right ]\in \mathbb{R}^{s \times K},
$$
where $s = \frac{r(r+1)}{2}$. Each vector $\widehat{\bq}^2_j$ is defined, according to \cite{Peherstorfer16DataDriven}, as
\begin{equation*}
\widehat{\bq}^2_j=
\begin{bmatrix}
\bq^{(1)}_j \\ \vdots \\ \bq^{(r)}_j
\end{bmatrix} \in \mathbb{R}^s,
\qquad \text{where}, \qquad
\bq^{(i)}_j = \widehat{q}_{i,j} \begin{bmatrix} \widehat{q}_{i,j} \\ \vdots \\ \widehat{q}_{r,j} \end{bmatrix} \in \mathbb{R}^i,
\end{equation*}
and $\widehat{q}_{i,j}$ is the $i$th element of the vector $\widehat{\bq}_j$. Now, instead of learning the operator $\widehat{\bH} \in \mathbb{R}^{r \times r^2}$, the operator $\widehat{\bF} \in \mathbb{R}^{r \times s}$ is learned, which satisfies the equivalent least squares problem
\begin{equation}
\min_{ \widehat{\bA} \in \mathbb{R}^{r \times r} , \widehat{\bF} \in \mathbb{R}^{r \times s}, \widehat{\bB}  \in \mathbb{R}^{r \times m}, \bhc \in \mathbb{R}^r}
\left \Vert   \widehat{\bQ}^{\top}\widehat{\bA}^{\top} + (\widehat{\bQ}^2)^{\top} \widehat{\bF}^{\top}  + \mathbf{U}^{\top}\widehat{\bB}^{\top} + \mathbf{1}_K \bhc^{\top}- \dot{\widehat{\bQ}}^{\top} \right \Vert^2_2
\label{eq:min4}
\end{equation}
The least squares problem is again of the form~\eqref{eq:min_OD} but we use the data matrix $\mathbf{D} = \begin{bmatrix} \widehat{\bQ}^{\top} \quad ( \widehat{\bQ}^2)^{\top} \quad  \mathbf{U}^{\top} \quad \mathbf{1}_K\end{bmatrix} \in \mathbb{R}^{K \times (r + s+ m+ 1)}$ and solve for the operators $\mathbf{O} = [\widehat{\bA} \quad \widehat{\bF} \quad \widehat{\bB} \quad \bhc] \in \mathbb{R}^{r \times (r+s+m+1)}$. Once we have solved for the operator $\widehat{\bF}$ we can easily transform it to obtain $\widehat{\bH}$.

\subsubsection{Regularization} \label{remark_reg}

We use an L$_2$-regularization (also known as Tikhonov regularization or ridge regression) to solve the operator inference problem, as shown in Eq.~\eqref{eq:min4}. The regularization term introduces a trade-off between operators that fit the data well and operators with small values. This regularization is used to avoid overfitting to the data, which in this setting causes our learned ROMs to be unstable ({\Rev solution blow up in finite time}). The regularization parameter $\lambda$ affects the performance of this algorithm---we require enough regularization to avoid overfitting, but if $\lambda$ is too large, the data will be poorly fit.

To help determine appropriate values of $\lambda$, we consider the ``L-curve" criterion discussed in~\cite{Lcurve}. The L-curve is a way of visualizing the effects of different values of $\lambda$ on the norm of the residual (data fit) against the norm of the solution. The L-curve criterion recommends choosing a value for $\lambda$ that lies in the corner of the curve, nearest the origin. In our numerical experiments, we compute the L-curve to help determine appropriate values for $\lambda$.

\begin{figure}[h!]
	\centering
	\includegraphics[width= 0.55\textwidth]{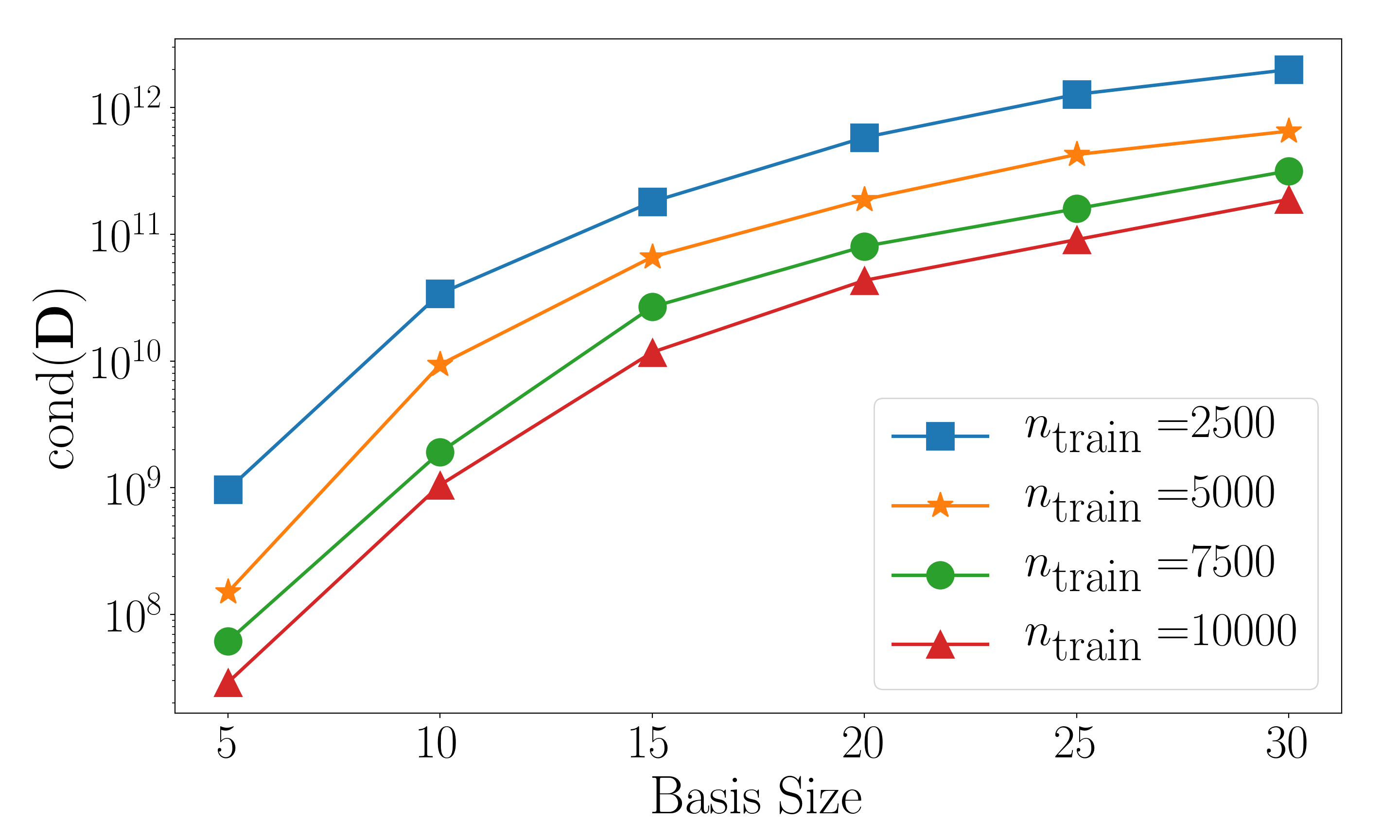}
	\caption{\label{fig:condition} The condition number of the data matrix, $\mathbf{D}$,  vs. basis size for different sized training sets.}
\end{figure}

Regularization also helps to reduce the condition number of the regularized least squares data matrix, $[\mathbf{D} \ \ \lambda \mathbf{P}]^\top$. In Figure~\ref{fig:condition} we show the condition number of the data matrix $\mathbf{D}$ from Eq.~\eqref{eq:min4} (with the data already scaled to [-1,1]) when we include $2500, 5000, 7500$ and $10000$ snapshots in the training set. The condition number is quite large for this application, yet it decreases as we add more training data. This effect is due to the fact that as we add more training data, the first $r$ POD basis vectors become richer. This in turn means that the projected data in $\mathbf{D}$ are richer for a given dimension of the POD basis. Our numerical experiments reinforced this finding by confirming that $10000$ training snapshots were required to achieve a sufficiently rich POD basis to obtain accurate ROMs, an indication of the complexity of the combustor dynamics.
{ \Rev While the condition number of the least-squares problem is high, for this example, it remains at manageable levels. However, if the condition number gets much larger, a careful treatment and consideration of proper data sampling and regularization will be required. 
}

\subsubsection{Error measures}
To evaluate the ROM performance, we define appropriate error metrics for each variable. The error is computed after the ROM solutions have been reconstructed back into the full CFD model dimension and scaled back to the original variable ranges. Recall Table~\ref{table:range_of_values1}, which showed the range of values for each variable. Below we provide details on how error is computed for each variable:
\begin{itemize}

\item For pressure and temperature, the values are always positive and well above zero, so we use a standard relative error, defined as
\begin{equation}
\mathcal{E}_{\textrm{relative},i} =  \frac{\vert \boldsymbol{\zeta}_{\textrm{CFD},i} - \boldsymbol{\zeta}_{\textrm{ROM},i}\vert}{\vert \boldsymbol{\zeta}_{\textrm{CFD},i} \vert}.
\label{eq:relative_error}
\end{equation}
where $\boldsymbol{\zeta} \in \{ \bp, \mathbf{T} \}$ and $\boldsymbol{\zeta}_{\cdot,i}$ denotes the $i$th entry of a vector, and $\boldsymbol{\zeta}_{\textrm{CFD}}$ is the CFD solution variable and $\boldsymbol{\zeta}_{\textrm{ROM}}$ is the solution variable obtained from the ROM simulation.

\item Due to the small values of species concentrations (on the order of $10^{-12}$), dividing by the true value can skew a small error. Similarly, velocities range from positive to negative, including zero. Thus, for species concentrations and velocities, we use a normalized absolute error, defined as
\begin{equation}
\mathcal{E}_{\textrm{nabs},i} = \frac{\vert \boldsymbol{\zeta}_{\textrm{CFD},i} - \boldsymbol{\zeta}_{\textrm{ROM},i}\vert}{\max_l \left  (\vert \boldsymbol{\zeta}_{\textrm{CFD},l}  \vert \right ) },
\label{eq:normalized_abs_error}
\end{equation}
where $\boldsymbol{\zeta} \in \{ \bv_x, \bv_y, [\mathbf{CH}_4], [\mathbf{O}_2], [\mathbf{CO}_2], [\bH_2\mathbf{O}] \}$ and $\max_{l}  \left ( \vert \boldsymbol{\zeta}_{\textrm{CFD,l}} \vert \right)$ denotes the maximum entry of $| \boldsymbol{\zeta}_{\textrm{CFD}}|$, i.e., the maximum absolute value over the discretized spatial domain.
\end{itemize}

\subsection{Learned reduced model performance} \label{sec:combustion_results}
The given $K=10000$ snapshots representing 1ms of GEMS simulation data {\Rev(scaled to [-1,1])} are used to learn the ROM. We are also given another {\Rev{2ms}} of testing data at the monitor locations shown in Figure~\ref{fig:combustor2}, which we use to assess the predictive capabilities of our learned ROMs beyond the range of training data.

\begin{figure}[h!]
	\centering
	\includegraphics[width = 0.55 \textwidth]{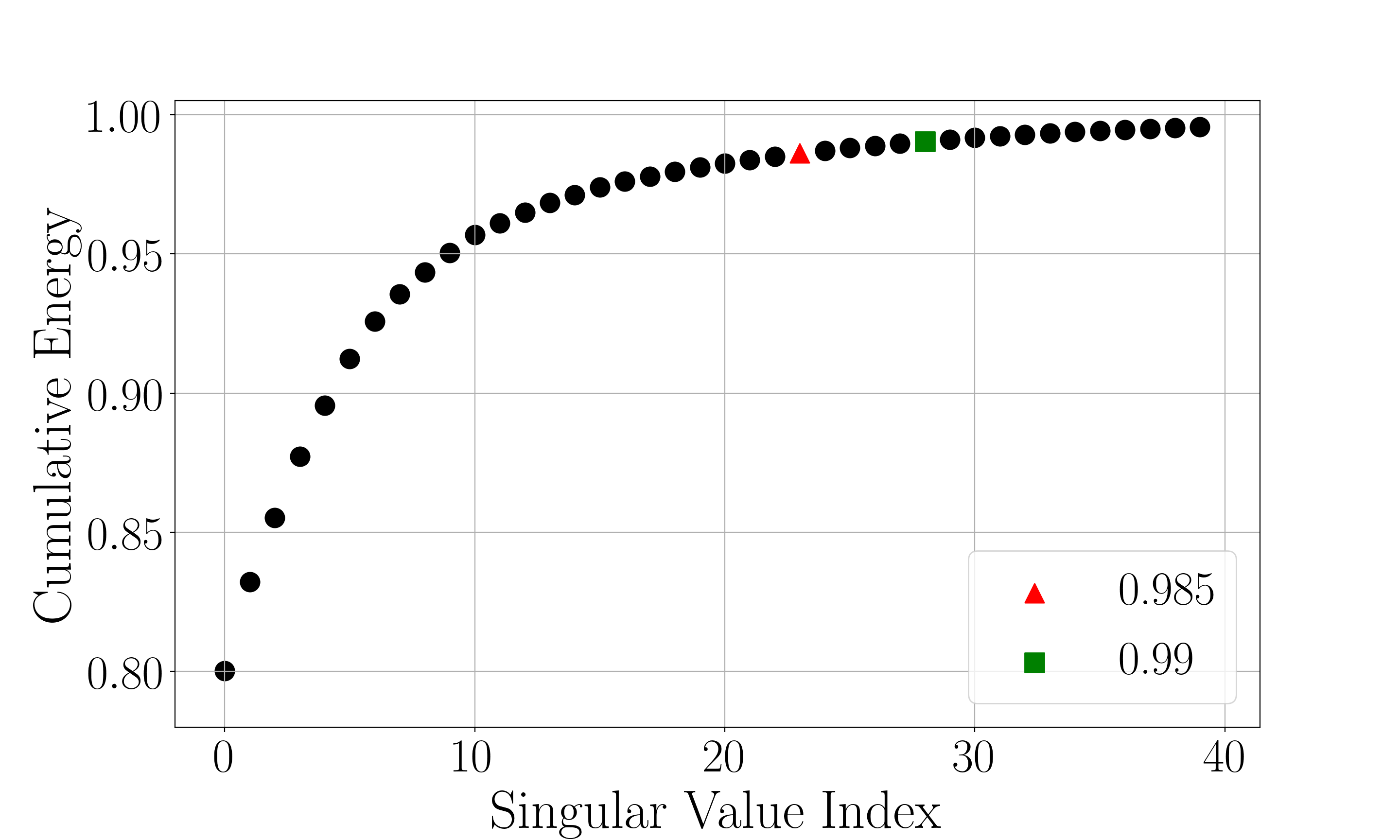}
	\caption{\label{fig:nt1000CumEn} The cumulative energy computed with Eq.~\eqref{eq:projection_error}. The leading {\Rev {$r = 24$ singular values capture 98.5\%}} of the energy and the leading $r = 29$ capture 99\%. }
\end{figure}

The cumulative energy of the singular values of the snapshot data matrix $\bQ$ is shown in Figure~\ref{fig:nt1000CumEn}. The singular values that correspond to a cumulative energy of 98.5\%  and 99\%  are indicated with the red triangle and green square, respectively. We use basis sizes of {\Rev{ $r=24$, capturing 98.5\%}} of the total energy, and $r = 29$, capturing 99\% of the total energy.

\begin{figure}
	\begin{subfigure}[t]{0.49\textwidth}
		\includegraphics[width = \textwidth,trim={1cm 0 1cm 0},clip]{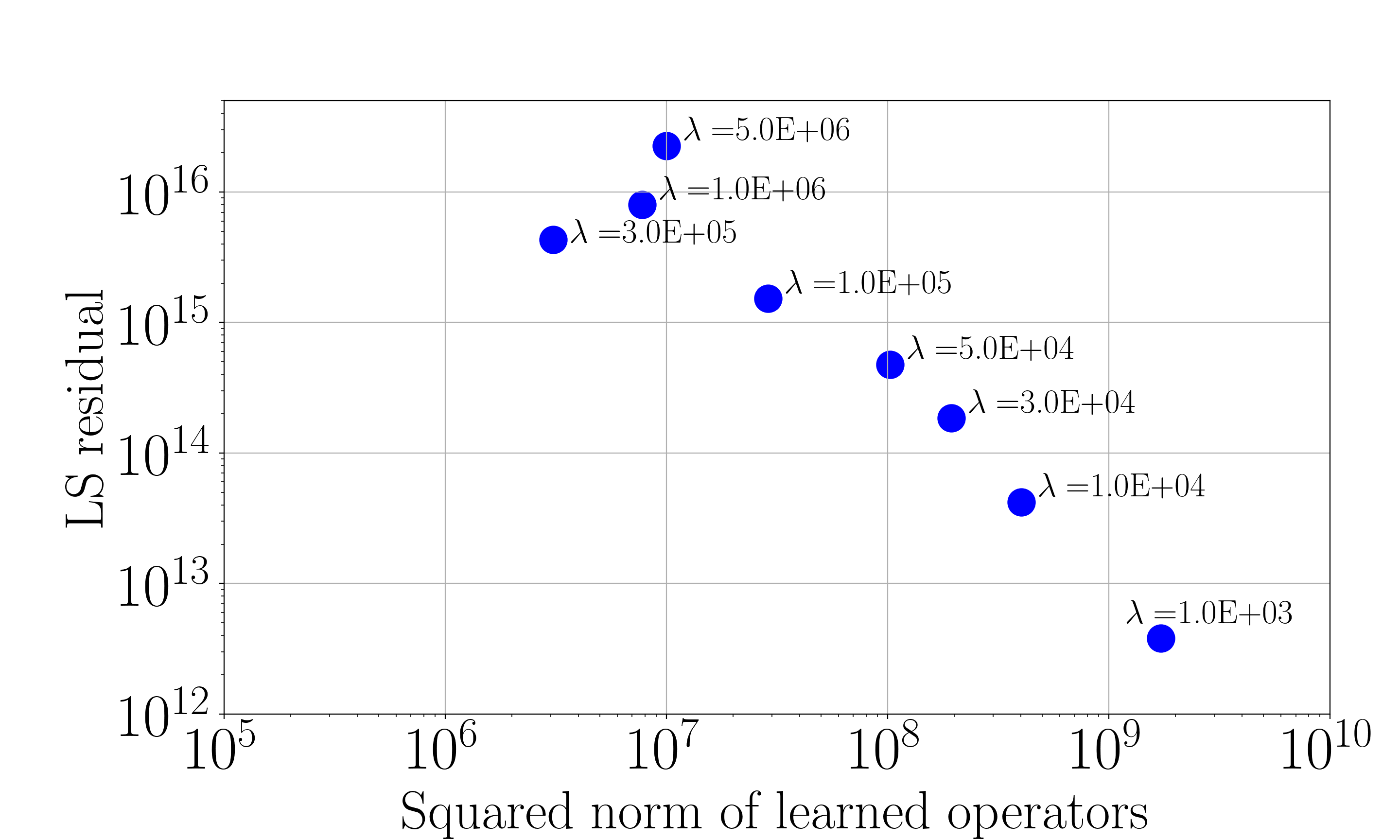}
		\caption{\label{fig:Lcurve_r24_nt10000} Basis size $r = 24$. }
	\end{subfigure}
	~\hfill
	\begin{subfigure}[t]{0.49\textwidth}
		\includegraphics[width = \textwidth,trim={1cm 0 1cm 0},clip]{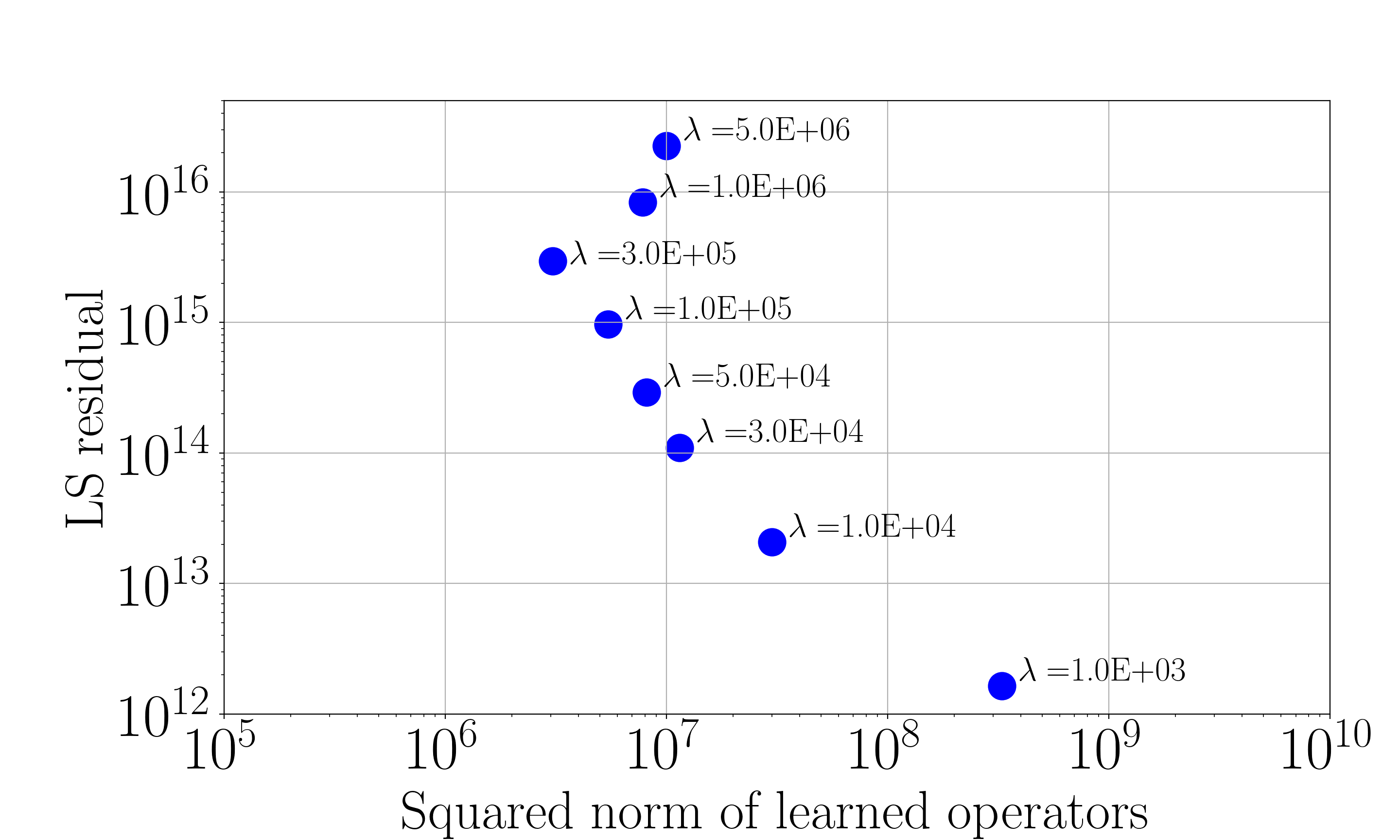}
		\caption{\label{fig:Lcurve_r29_nt10000} Basis size $r = 29$.}
	\end{subfigure}
	\caption{\label{fig:nt1000Lcurve}The L-curve for different basis sizes and regularization parameters $\lambda$. The horizontal axis shows the squared norm of learned operators, $\Vert \mathbf{O}\Vert_2^2$, and the vertical axis shows the least squares residual,  $\Vert \mathbf{D} \mathbf{O}^{\top} - \dot{\widehat{\bQ}}^{\top} \Vert^2_2$.}
\end{figure}

{\Rev In Figures~\ref{fig:Lcurve_r24_nt10000},~\ref{fig:Lcurve_r29_nt10000} we show the L-curve for each basis size. The L-curve for a basis of $r=24$ is somewhat uninformative in this case. The regularization parameters chosen were $1.0$E+05 and $\lambda = 3.0$E+05, which lie in the upper part of the L-curve and are therefore a recommended choice by the L-curve criterion discussed above. For a basis of size of $r=29$, the L-curve indicates a regularization parameter around $\lambda = 3.0$E+04. Stable systems are produced for $\lambda = 3.0$E+04 and $5.0$E+04. }

\begin{figure}[h!]
\centering
\begin{subfigure}[t]{0.49\textwidth}
\includegraphics[width = \textwidth]{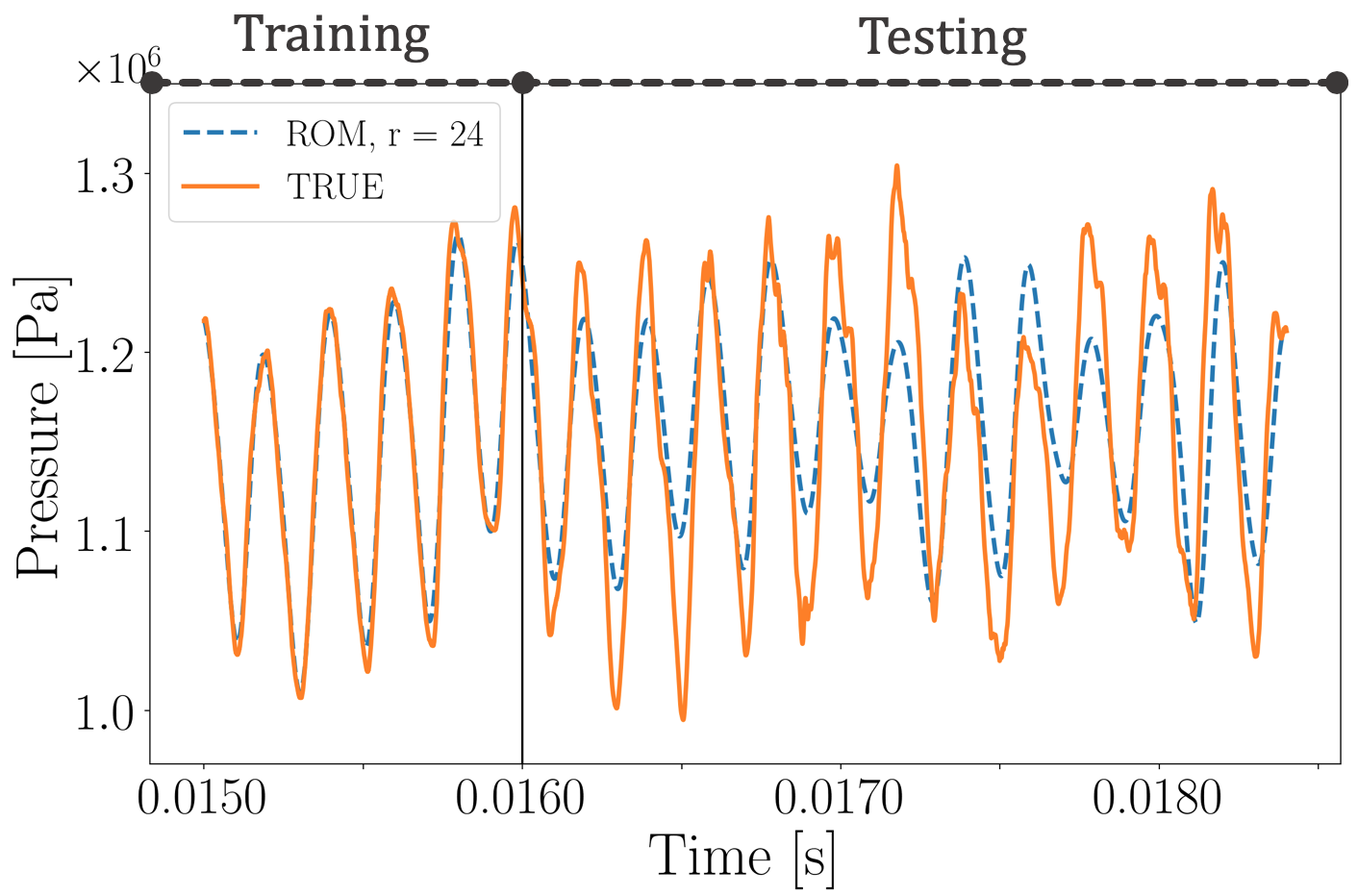}
\caption{\label{fig:tt_r24_1} $\lambda = 1.0\rm{E}+05$. }
\end{subfigure}
~\hfill
\begin{subfigure}[t]{0.49\textwidth}
\includegraphics[width = \textwidth,trim={0 0.1cm 0 0},clip]{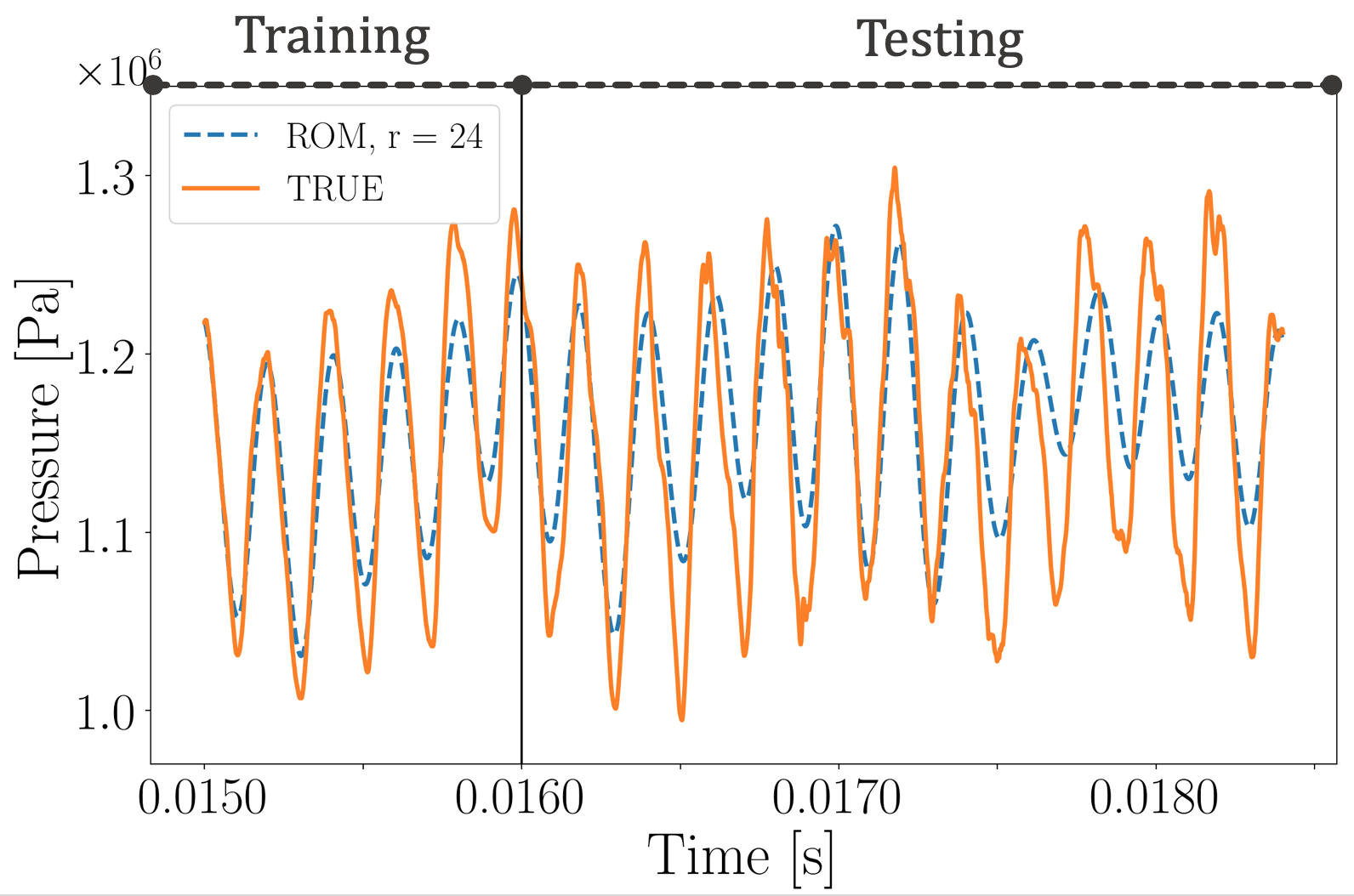}
\caption{\label{fig:tt_r24_2} $\lambda = 3.0\rm{E}+05$.}
\end{subfigure}
\caption{\label{fig:timetraces_r24}Pressure time traces for basis size {\Rev{$r = 24$}}. Training with 10000 snapshots. Black vertical line denotes the end of the training data and the beginning of the test data.}
\end{figure}

\begin{figure}[h!]
\centering
\begin{subfigure}[t]{0.49\textwidth}
\includegraphics[width=\textwidth]{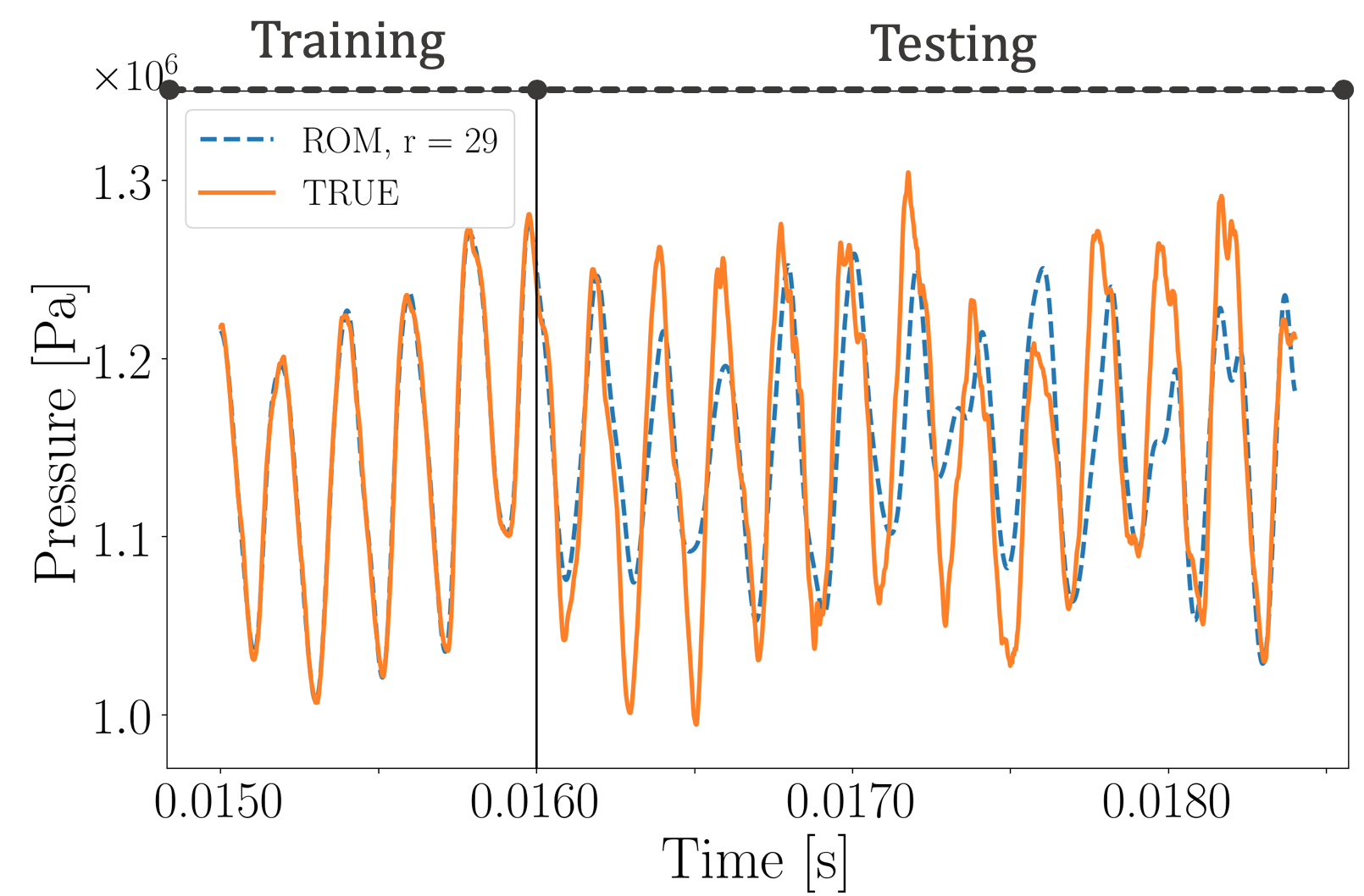}
\caption{\label{fig:tt_r29_1} $\lambda = 3.0\rm{E}+04$. }
\end{subfigure}
~\hfill
\begin{subfigure}[t]{0.49\textwidth}
\includegraphics[width=\textwidth,trim={0 0.1cm 0 0},clip]{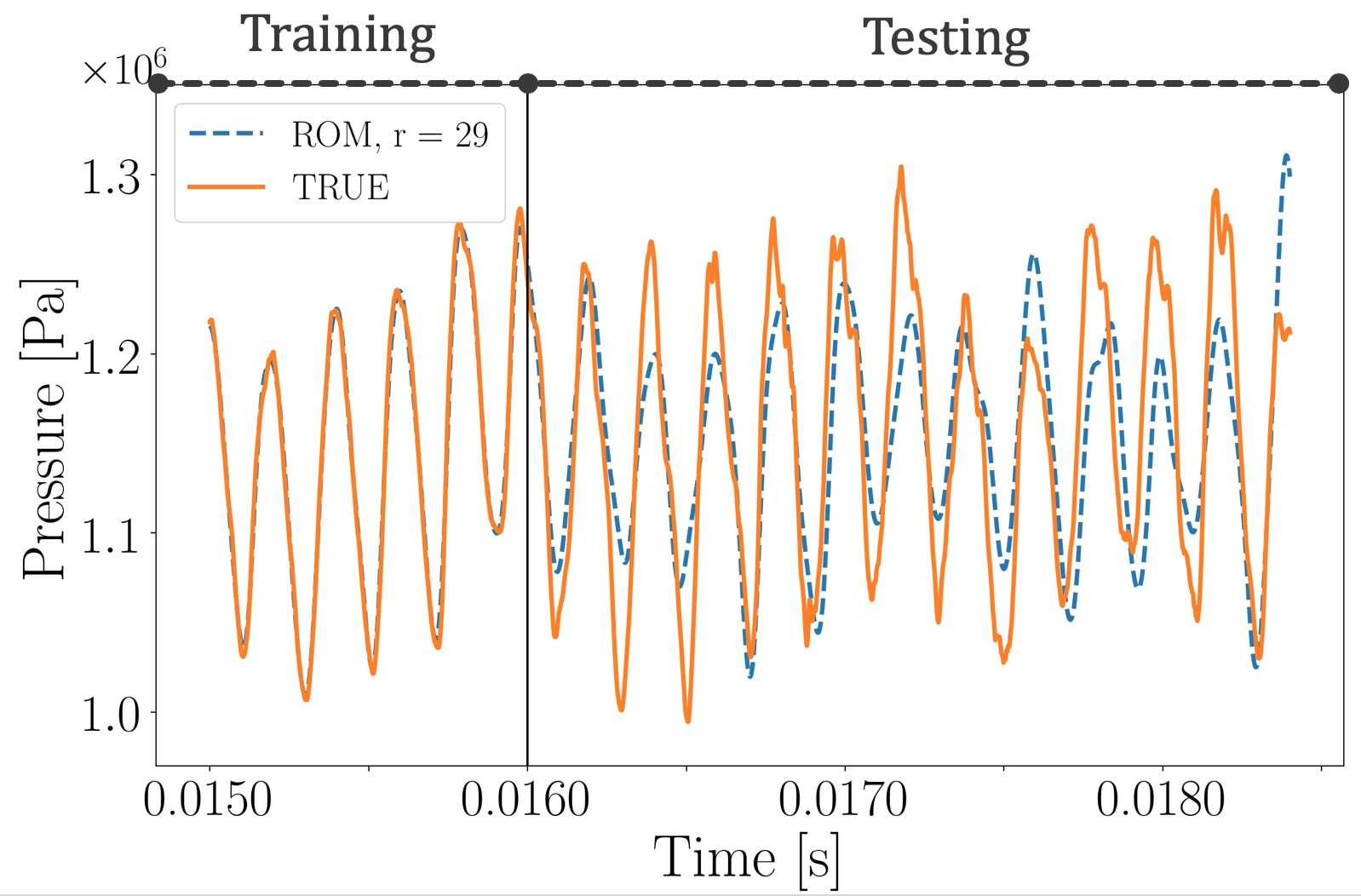}
\caption{\label{fig:tt_r29_2} $\lambda = 5.0\rm{E}+04$.}
\end{subfigure}
\caption{\label{fig:timetraces_r29}Pressure time traces for basis size of $r = 29$. Training with 10000 snapshots. Black vertical line denotes the end of the training data and the beginning of the test data.}
\end{figure}

We simulate the learned ROM for the two model sizes {\Rev{ $r=24$}} and $r=29$ with the same initial value and time step size ($\Delta t = 1\times 10^{-7}$s) as those used to generate the training set. {\Rev Since the ROM was constructed from data scaled to [-1,1], we solve the ROM in the (scaled) subspace, and then reconstruct the dimensional quantities by reversing the scaling.}
Figures~\ref{fig:timetraces_r24} and \ref{fig:timetraces_r29} compare the time trace of pressure computed by GEMS (our ``truth'' data) with the ROM predictions for 30000 time steps {\Rev {(10000 time steps used for training, 20000 time steps are pure prediction for the ROM)}} at the cell located at $(0.0,0.0225)$ in the domain (denoted as monitor location 1 in Figure~\ref{fig:combustor2}). The performance of the ROM on the training data (first 1ms of data) is good in both cases, although the $r=29$ ROM (Figure~\ref{fig:timetraces_r29}) is more accurate. {\Rev{For test data predictions beyond the training data, both ROMs yield accurate phase predictions and pressure oscillation amplitudes that are good approximations of the truth.}}

{\Rev A Galerkin-projection-based POD method was applied to this GEMS model in Ref.~\cite{cheng}. The authors there found that a large number of modes ($r > 100$) was necessary to obtain stable ROMs. Moreover, comparable accuracy, e.g., for the pressure probe predictions shown here,  was only achieved with $r = 200$ POD modes for the ROM (c.f. Fig.~15 in Ref.~\cite{cheng} with Figures~\ref{fig:timetraces_r24}--\ref{fig:timetraces_r29} herein; both are recording pressure at the same probe location). Consequently, while our learned ROM does not resolve the full  flow physics, we do obtain good predictability in time at much lower ROM dimension than in the classical Galerkin-POD approach used in Ref.~\cite{cheng}. 
The key innovation leading to this improvement is our use of variable transformations to build the ROM over a space for which the transformed governing equations have more polynomial structure. The non-intrusive operator inference approach is an enabler that makes these variable transformations practical from an implementation perspective, since the transformations are applied only to the snapshot data and not to the CFD model itself.
}

\begin{figure}[h!]
\centering
\includegraphics[width  = 0.48\textwidth]{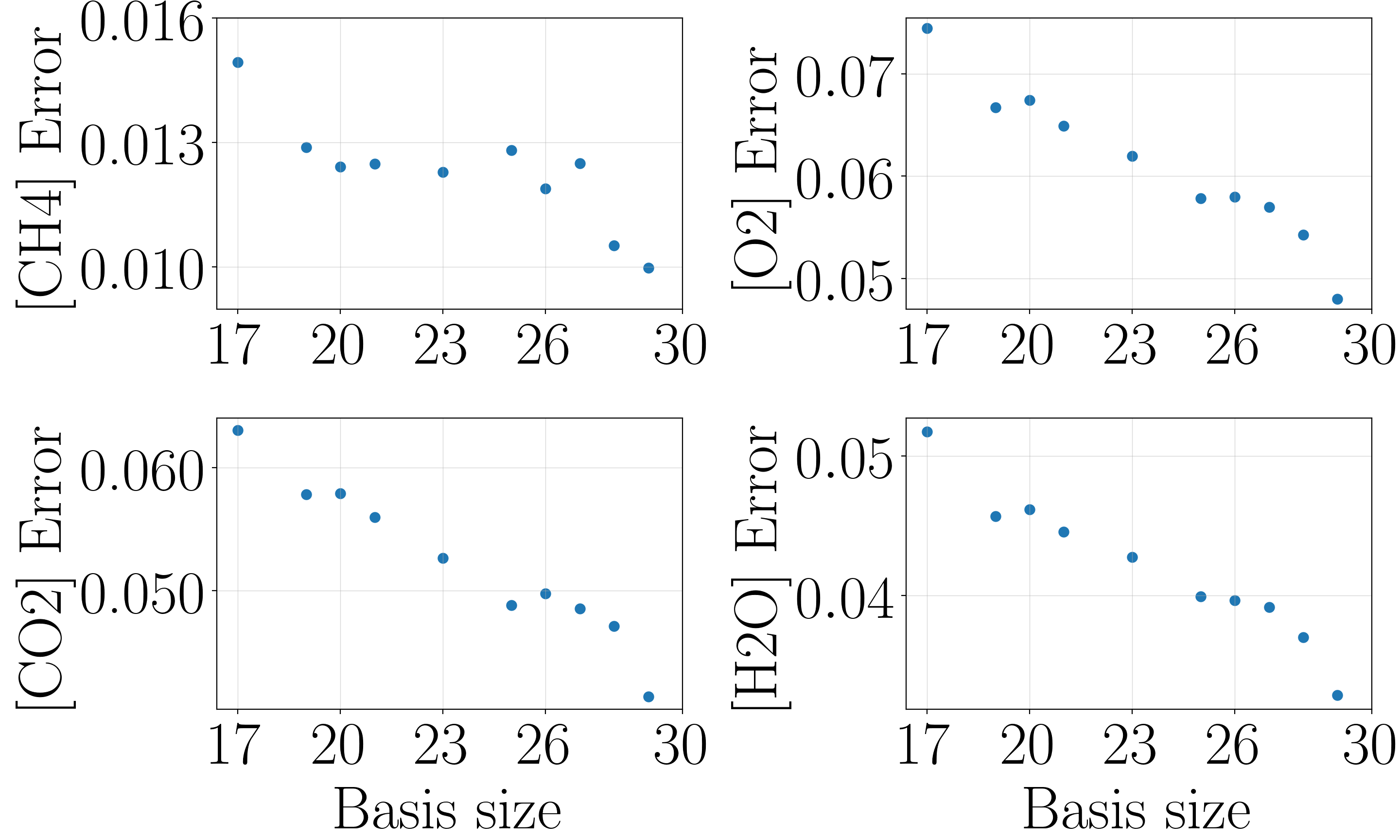}
\hspace{0.02\textwidth}
\includegraphics[width = 0.49\textwidth]{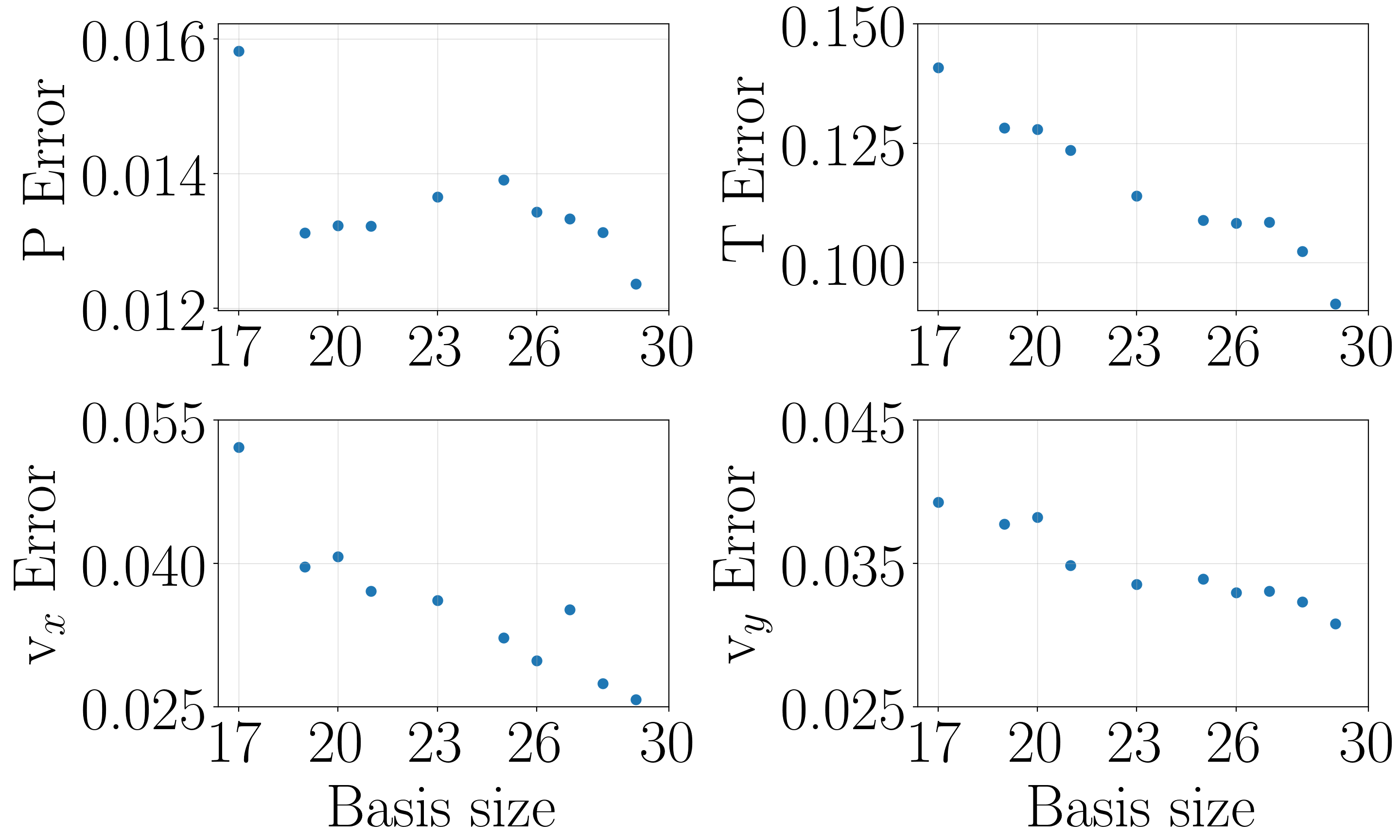}
\caption{\label{fig:avg_error1} Error measures vs.\ basis size, averaged over the spatial domain at the last time step of training data. Normalized absolute error~\eqref{eq:normalized_abs_error} of species $\text{CH}_4, \text{O}_2, \text{CO}_2, \text{H}_2\text{O}$ and $\bv_x$ and $\bv_y$ velocities; Relative error~\eqref{eq:relative_error} given for pressure and temperature. }
\end{figure}

We also compute the average error of each variable over the entire domain at the last time step of the training set (the 10000th time step). The normalized absolute error, defined in Eq.~\eqref{eq:normalized_abs_error}, is shown for each species and for $x$ and $y$ velocity in Figure~\ref{fig:avg_error1}. This figure also shows the relative error, defined in Eq.~\eqref{eq:relative_error}, for pressure and temperature. These plots show that overall, the error is decreasing with an increasing basis size. At a basis size of 18, 22, and 24 the systems were unstable ({\Rev solution blow up in finite time}) and so these basis sizes are excluded from the figure. The cause of this{\Rev, and the non-monotone error decay,} may be due to the fact that the same regularization parameter was used for each of these, $\lambda = 3.0$E+04. Ideally, one would pick a parameter specific for the basis size{\Rev, but here we used the same regularization parameter in order to give a fairer assessment of the ROM performance without using manual tuning to optimize our results. We note, however, that even with manual tuning of the regularization parameters we are not guaranteed monotone state-error decay for strongly nonlinear dynamical system ROMs.}

\begin{figure}[h!]
\centering
\includegraphics[width  = 0.7\textwidth]{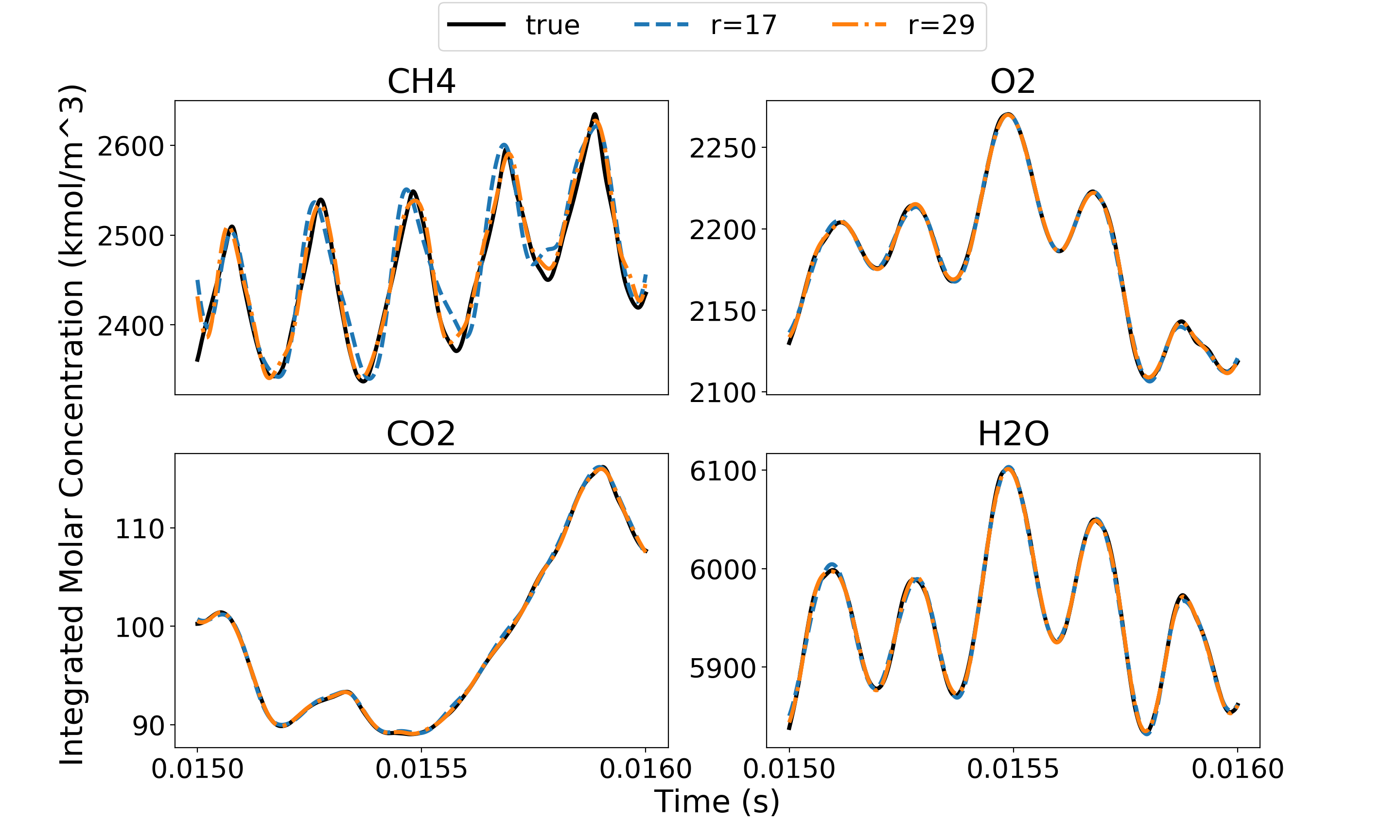}
\caption{\label{fig:int_species} Integrated species at each time step for different basis sizes. Training with 10000 snapshots.}
\end{figure}

In Figure~\ref{fig:int_species}, we show the integrated species concentrations over time. To compute these, at each time step in our simulation, we sum all elements of a species vector. This measure monitors whether our ROM conserves species mass, an important feature of a physically meaningful simulation. As the discretization of the high-fidelity model becomes finer, point-wise error may become large and misleading if the mass is shifted slightly into the neighboring cells. The integrated species concentration complements the evaluation of point-wise errors and provides a global view of the error in the domain. {\Rev While $\text{CH}_4$ conservation is tracked well qualitatively by the ROM, it does show the largest deviation out of the four species. This is a result of $\text{CH}_4$ having the sharpest gradients ($\text{CH}_4$ concentration ranging from 0 to 1) compared to the other three species, see also Figures~\ref{fig:CH4}--\ref{fig:H2O} below, and their respective color bars.}

\begin{figure}[h!]
\begin{subfigure}[t]{0.49\textwidth}
\includegraphics[width = \textwidth,trim={0 .1cm .1cm 0},clip]{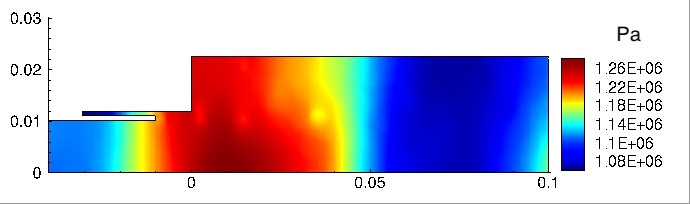}
\caption{\label{fig:true_P} True pressure. }
\end{subfigure}
~\hfill
\begin{subfigure}[t]{0.49\textwidth}
	\includegraphics[width = \textwidth,trim={0 .1cm .1cm 0},clip]{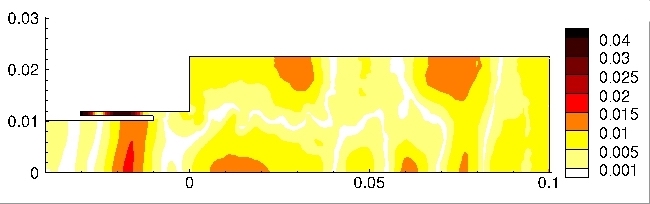}
	\caption{\label{fig:error_P} Relative error of pressure.}
\end{subfigure}
~\vfill
\begin{subfigure}[t]{0.49\textwidth}
	\includegraphics[width = \textwidth,trim={0 .1cm .1cm 0},clip]{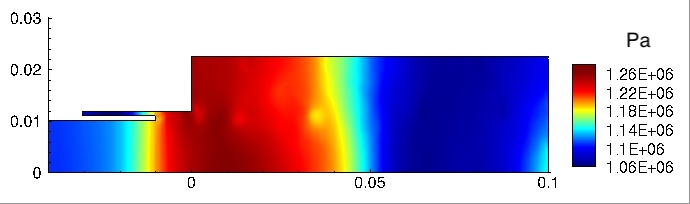}
	\caption{\label{fig:r29_P} Predicted pressure.}
\end{subfigure}
\caption{\label{fig:pressure}Predictive results for pressure at the last time step of training data. Training with 10000 snapshots, a basis size of $r=29$, and regularization set to $\lambda = 3.0\rm{E}+04$.}
\end{figure}

\begin{figure}[h!]
\begin{subfigure}[t]{0.49\textwidth}
\includegraphics[width = \textwidth,trim={0.1cm .5cm .1cm 0.1cm },clip]{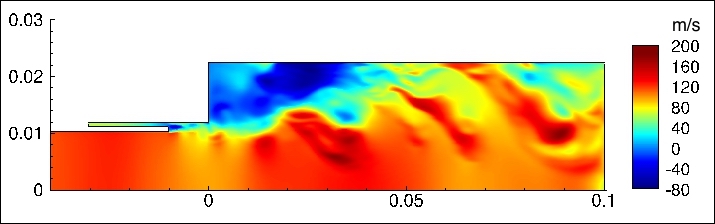}
\caption{\label{fig:true_U} True velocity $\bv_x$.}
\end{subfigure}
~\hfill
\begin{subfigure}[t]{0.49\textwidth}
	\includegraphics[width = \textwidth,trim={0.1cm .5cm .1cm 0.1cm},clip]{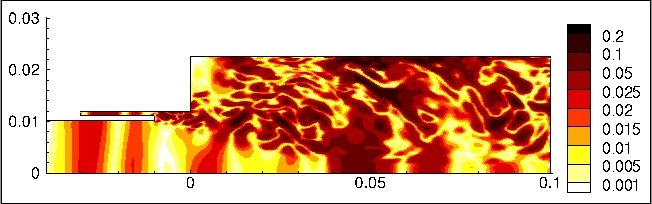}
	\caption{\label{fig:error_U} Normalized absolute error of velocity $\bv_x$.}
\end{subfigure}
~\vfill
\begin{subfigure}[t]{0.49\textwidth}
	\includegraphics[width = \textwidth,trim={0.1cm .5cm .1cm 0.1cm},clip]{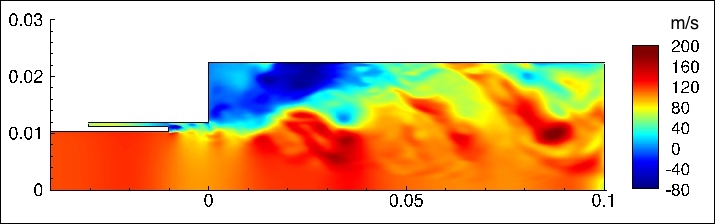}
	\caption{\label{fig:r29_U} ROM-predicted velocity $\bv_x$.}
\end{subfigure}
\caption{\label{fig:xvelocity}Predictive results for velocity $\bv_x$ at the last time step of training data. Training with 10000 snapshots, a basis size of $r=29$, and regularization set to $\lambda = 3.0\rm{E}+04$.}
\end{figure}

\begin{figure}[h!]
\begin{subfigure}[t]{0.49\textwidth}
\includegraphics[width = \textwidth,trim={0 .1cm .1cm 0},clip]{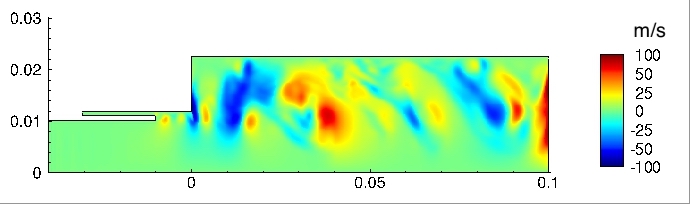}
\caption{\label{fig:true_V} True velocity $\bv_y$.}
\end{subfigure}
~\hfill
\begin{subfigure}[t]{0.49\textwidth}
	\includegraphics[width = \textwidth,trim={0.1cm .5cm .1cm 0.1cm},clip]{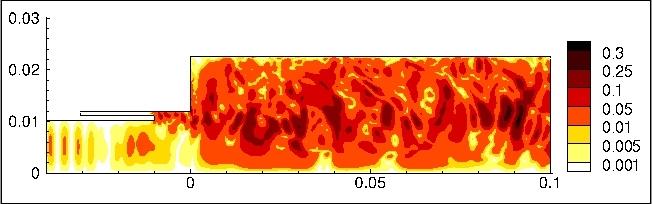}
	\caption{\label{fig:error_V} Normalized absolute error of velocity $\bv_y$.}
\end{subfigure}
~\vfill
\begin{subfigure}[t]{0.49\textwidth}
	\includegraphics[width = \textwidth,trim={0 .1cm .1cm 0},clip]{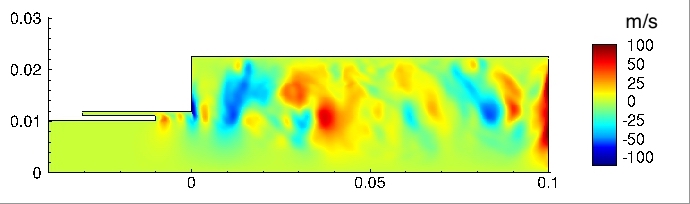}
	\caption{\label{fig:r29_V} ROM-predicted velocity $\bv_y$.}
\end{subfigure}
\caption{\label{fig:yvelocity}Predictive results for velocity $\bv_y$ at the last time step of training data. Training with 10000 snapshots, a basis size of $r=29$, and regularization set to $\lambda = 3.0\rm{E}+04$.}
\end{figure}

\begin{figure}[h!]
\begin{subfigure}[t]{0.49\textwidth}
\includegraphics[width = \textwidth,trim={0 .1cm .1cm 0},clip]{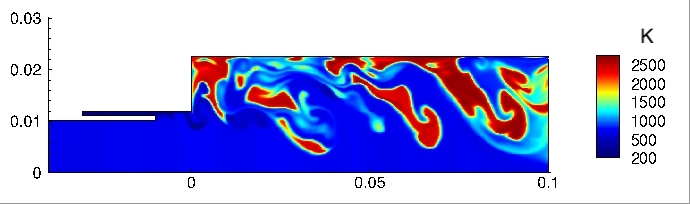}
\caption{\label{fig:true_T} True temperature.}
\end{subfigure}
~\hfill
\begin{subfigure}[t]{0.49\textwidth}
	\includegraphics[width = \textwidth,trim={0 .1cm .1cm 0},clip]{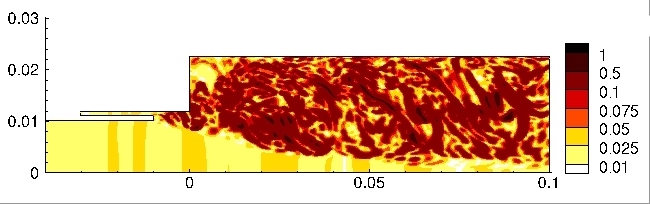}
	\caption{\label{fig:error_T} Relative error of temperature.}
\end{subfigure}
~\vfill
\begin{subfigure}[t]{0.49\textwidth}
	\includegraphics[width = \textwidth,trim={0 .1cm .1cm 0},clip]{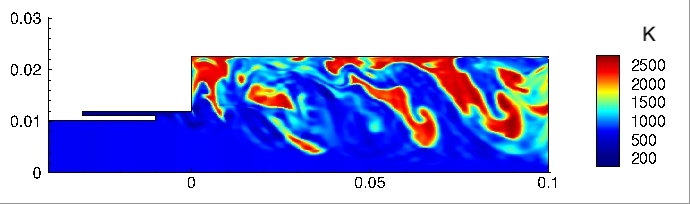}
	\caption{\label{fig:r29_T} Predicted temperature.}
\end{subfigure}
\caption{\label{fig:temperature}Predictive results for temperature at the last time step of training data. Training with 10000 snapshots, a basis size of $r=29$, and regularization set to $\lambda = 3.0\rm{E}+04$.}
\end{figure}

We also compare the state variables over the entire domain predicted by the learned ROM with the given GEMS data at the last time step $K=10000$ (which corresponds to $t=0.0159999$s). We provide the true field, the ROM-predicted field, and an error field for each variable in Figures~\ref{fig:pressure}--\ref{fig:H2O}. Again, for pressure and temperature, we use a relative error from Eq.~\eqref{eq:relative_error}. For $x$ and $y$ velocity and for species molar concentrations we use a normalized absolute error from Eq.~\eqref{eq:normalized_abs_error}. The plots show that the ROM predictions are, as expected, not perfect, but indeed they capture well the overall structure and many details of the solution fields.

\begin{figure}[h!]
\begin{subfigure}[t]{0.49\textwidth}
\includegraphics[width = \textwidth,trim={0 .1cm .1cm 0},clip]{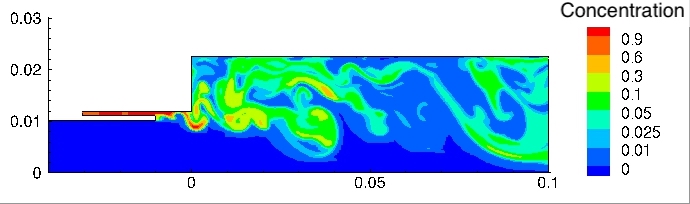}
\caption{\label{fig:true_CH4} True CH$_4$. }
\end{subfigure}
~\hfill
\begin{subfigure}[t]{0.49\textwidth}
	\includegraphics[width = \textwidth,trim={0 .1cm .1cm 0},clip]{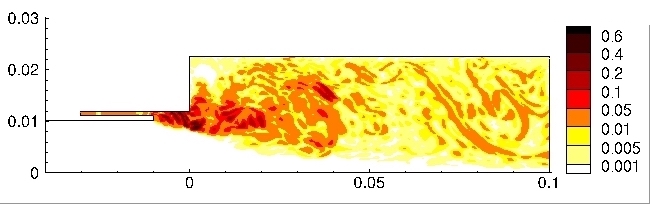}
	\caption{\label{fig:error_CH4} Normalized absolute error of CH$_4$. }
\end{subfigure}
~\vfill
\begin{subfigure}[t]{0.49\textwidth}
	\includegraphics[width = \textwidth,trim={0 .1cm .1cm 0},clip]{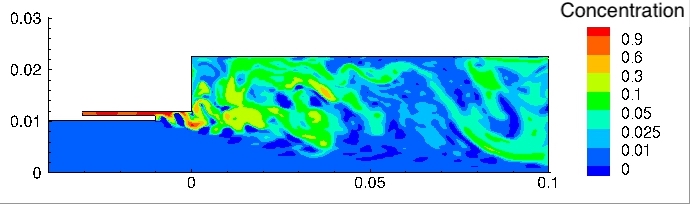}
	\caption{\label{fig:r29_CH4} Predicted CH$_4$.}
\end{subfigure}
\caption{\label{fig:CH4} Predictive results for CH$_4$ molar concentration at the last time step of training data. Training with 10000 snapshots, a basis size of $r=29$, and regularization set to $\lambda = 3.0\rm{E}+04$.}
\end{figure}

\begin{figure}[h!]
\begin{subfigure}[t]{0.49\textwidth}
\includegraphics[width = \textwidth,trim={0 .1cm .1cm 0},clip]{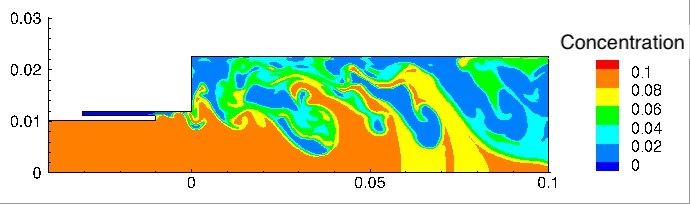}
\caption{\label{fig:true_O2} True O$_2$. }
\end{subfigure}
~\hfill
\begin{subfigure}[t]{0.49\textwidth}
	\includegraphics[width = \textwidth,trim={0 .1cm .1cm 0},clip]{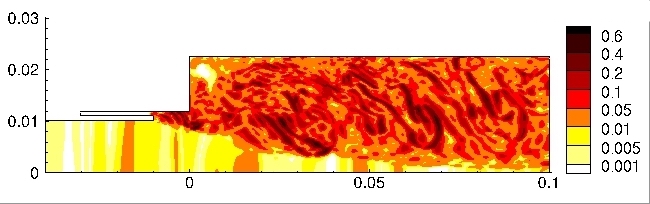}
	\caption{\label{fig:error_O2} Normalized absolute error of O$_2$.}
\end{subfigure}
~\vfill
\begin{subfigure}[t]{0.49\textwidth}
	\includegraphics[width = \textwidth,trim={0 .1cm .1cm 0},clip]{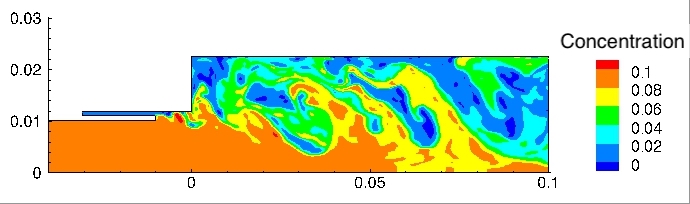}
	\caption{\label{fig:r29_O2} Predicted O$_2$. }
\end{subfigure}
\caption{\label{fig:O2} Predictive results for O$_2$ molar concentration at the last time step of training data. Training with 10000 snapshots, a basis size of $r=29$, and regularization set to $\lambda = 3.0\rm{E}+04$.}
\end{figure}

\begin{figure}[h!]
\begin{subfigure}[t]{0.49\textwidth}
\includegraphics[width = \textwidth,trim={0 .1cm .1cm 0},clip]{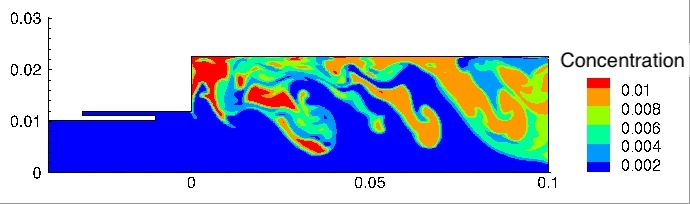}
\caption{\label{fig:true_CO2} True CO$_2$. }
\end{subfigure}
~\hfill
\begin{subfigure}[t]{0.49\textwidth}
	\includegraphics[width = \textwidth,trim={0 .1cm .1cm 0},clip]{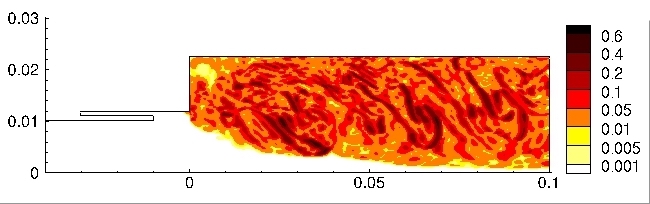}
	\caption{\label{fig:error_CO2} Normalized absolute error of CO$_2$. }
\end{subfigure}
~\vfill
\begin{subfigure}[t]{0.49\textwidth}
	\includegraphics[width = \textwidth,trim={0 .1cm .1cm 0},clip]{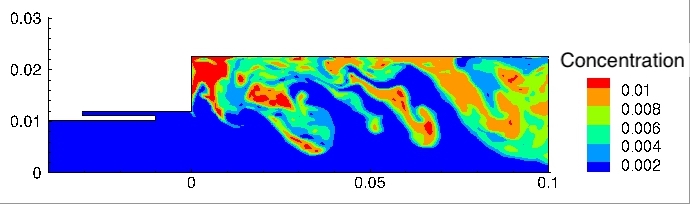}
	\caption{\label{fig:r29_CO2} Predicted CO$_2$.}
\end{subfigure}
\caption{\label{fig:CO2} Predictive results for CO$_2$ molar concentration at the last time step of training data. Training with 10000 snapshots, a basis size of $r=29$, and regularization set to $\lambda = 3.0\rm{E}+04$.}
\end{figure}

\begin{figure}[h!]
\begin{subfigure}[t]{0.49\textwidth}
\includegraphics[width = \textwidth,trim={0 .1cm .1cm 0},clip]{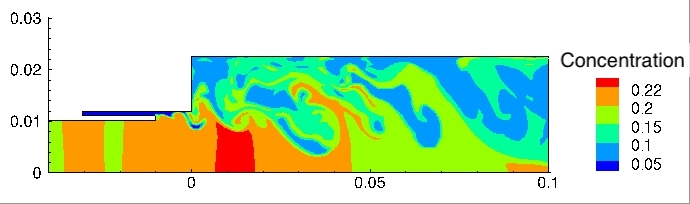}
\caption{\label{fig:true_H2O} True H$_2$O. }
\end{subfigure}
~\hfill
\begin{subfigure}[t]{0.49\textwidth}
	\includegraphics[width = \textwidth,trim={0 .1cm .1cm 0},clip]{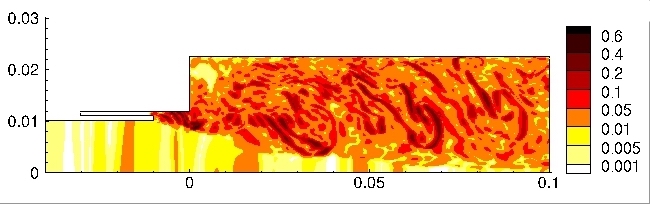}
	\caption{\label{fig:error_H2O} Normalized absolute error of H$_2$O. }
\end{subfigure}
~\vfill
\begin{subfigure}[t]{0.49\textwidth}
	\includegraphics[width = \textwidth,trim={0 .1cm .1cm 0},clip]{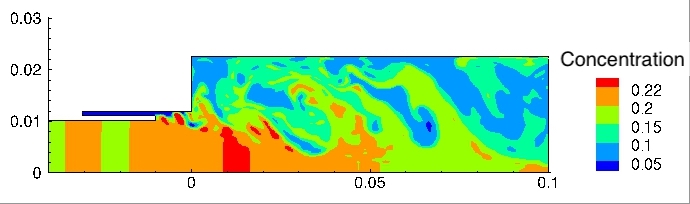}
	\caption{\label{fig:r29_H2O} Predicted H$_2$O. }
\end{subfigure}
\caption{\label{fig:H2O} Predictive results for H$_2$O molar concentration at the last time step of training data. Training with 10000 snapshots, a basis size of $r=29$, and regularization set to $\lambda = 3.0\rm{E}+04$.}
\end{figure}

Table~\ref{tbl:timing} shows timing results for the ROM generation and simulation, as performed using python 3.6.4 on a dual-core Intel i5 processor with 2.3 GHz and 16GB RAM. We report the following CPU times: solving the operator inference least squares problem~\eqref{eq:min4}; ROM runtime for the two different basis sizes for {\Rev3ms} of real time prediction; and reconstruction of the high-dimensional, unscaled combustion variables. The latter is required since the ROM is evolving the dynamics in the $[-1,1]$ scaled variables, and thus a post-processing step is required to obtain the true magnitudes of the variables.  In comparison with the approximately 200h of CPU time needed on a 40-core architecture (more details in Section~\ref{sec:GEMSdata_results}) to compute the first 10000 snapshots of data, the ROMs provide five to six orders of magnitude in computational speedup.

\begin{table}[h!]
	\begin{center}
		\caption{CPU times for two ROMs with time step size $\Delta t = 1\times 10^{-7}\text{s}$ and 30000 time steps.}
		\label{tbl:timing}
		\begin{tabular}{ c | r | r | r  }
			ROM order& LS solve of~\eqref{eq:final_minimization}& ROM simulation & Reconstructing high-dim. field\\
			\hline
			$r=24$ & 2.80\text{s} & 6.06\text{s}  & 0.04\text{s} \\
			$r=29$ & 6.05\text{s} & 6.22\text{s}  & 0.05\text{s}\\
		\end{tabular}%
	\end{center}
\end{table}

\section{Conclusion} \label{sec:conclusion}
Operator inference is a data-driven method for learning reduced-order models (ROMs) of dynamical systems with polynomial structure. This paper demonstrates how variable transformations can expose quadratic structure in the nonlinear system of partial differential equations describing a complex combustion model. This quadratic structure is preserved under projection, providing the mathematical justification for learning a quadratic ROM using operator inference. An important feature of the approach is that the learning of the ROM is non-intrusive---it requires state solutions generated by running the high-fidelity combustion model, but it does not require access to the discretized operators of the governing equations. This is important because it means that the variable transformations can be applied as a post-processing step to the simulation data set, but no intrusive modifications are needed to the high-fidelity code. While the quadratic model form is an approximation for this particular application problem, the numerical results show that the learned quadratic ROM can predict relevant quantities of interest and can also conserve species accurately. Many nonlinear equations in scientific and engineering applications admit variable transformations that expose polynomial structure. This combined with the non-intrusive nature of the approach make it a viable option for deriving ROMs for complex nonlinear applications where traditional intrusive model reduction is impractical and/or unreliable.
{\Rev While we improved ROM stability through the presented regularization of the least-squares problem, some of the ROMs were unstable. Future work thus includes devising alternative approaches (see e.g., \cite{kalashnikova2011stable,balajewicz2016minimal}) and developing theory to guarantee stable learned ROMs.}
%

\section*{Appendix A: Lifting chemical source terms} \label{app:S}
The chemical source terms in $\vec{S}$ in Eq.~\eqref{eq:defS} are $\dot{\omega}_l = M_l \dot{c}_l^{\text{reaction}}$, see Eq.~\eqref{eq:omega_dot}.  The dynamics for the source terms $\dot{c}_l^{\text{reaction}}$ are given by
\begin{subequations}
	\begin{align}
	\dot{c}_1^{\text{reaction}}  & = -A \exp \left (-\frac{E_a}{R_u T} \right ) \ \CHfour^{0.2}\Otwo^{1.3},\\
	\dot{c}_2^{\text{reaction}}  & =  2  \dot{c}_1^{\text{reaction}},\\
	\dot{c}_3^{\text{reaction}}  & = -\dot{c}_1^{\text{reaction}}, \\
	\dot{c}_4^{\text{reaction}}  & = -2 \dot{c}_1^{\text{reaction}}.
	\end{align}
	\label{eq:origChemistry}
\end{subequations}
To lift this system to polynomial form, we introduce the auxiliary variables
\begin{equation*}
w_1  = \CHfour^{0.2}, \quad  w_2 = \CHfour^{-1}, \quad w_3  =\Otwo^{1.3},\quad  w_4 =\Otwo^{-1}, \quad w_5  = \exp \left (-\frac{E_a}{R_u T} \right ), \quad w_{6} = \frac{1}{T^2}.
\end{equation*}
The source term dynamics~\eqref{eq:origChemistry} are then cubic in $w_1, w_3, w_5$:
\begin{subequations}
	\begin{align}
	\dot{c}_1^{\text{reaction}} & = -A w_1 w_3 w_5, 	\\
	\dot{c}_2^{\text{reaction}} & = 2 A w_1 w_3 w_5,\\
	\dot{c}_3^{\text{reaction}} & = -A w_1 w_3 w_5, \\
	\dot{c}_4^{\text{reaction}} & = -2 A w_1 w_3 w_5.
	\end{align}
	\label{eq:chemistry}
\end{subequations}
We next derive the dynamics for the auxiliary variables $w_1, \ldots, w_6$. For instance, we have $\dot{w}_1 = 0.2 \CHfour^{0.2-1} \dot{\CHfour} = 0.2 w_1  w_2 \dot{\CHfour}$. Similarly, we obtain the system of auxiliary dynamics:
\begin{align*}
\dot{w}_1 & =  0.2 w_1  w_2 \dot{\CHfour},  & \dot{w}_2 & =  -w_1^2 \dot{\CHfour}, \\
\dot{w}_3 & = 1.3 w_3 w_4\dot{\Otwo},  & \dot{w}_4 & = -w_4^2 \dot{\Otwo}, \\
\dot{w}_5 & = \frac{E_a}{R_u} \frac{1}{T^2} w_5 \dot{T} = \frac{E_a}{R_u} w_5 w_{6} \dot{T}, & \dot{w}_6 & = -2 \frac{1}{T^3} = -2 T w_{6}^2.  
\end{align*}	
The dynamics of the lifted variables are quintic in the variables $w_1, \ldots, w_6$.
If we further include an additional auxiliary variable $w_7 = w_1 w_3 w_5$, then we obtain the system of equations
\begin{align*}
\dot{w}_1 & =  0.2 A   w_1  w_2 w_7,   & \dot{w}_2 & = - A   w_1^2 w_7,  \\
\dot{w}_3 & =  2.6 A w_3 w_4 w_7, 	  & \dot{w}_4 & = -2 A w_4^2 w_7, \\
\dot{w}_5 & = \frac{E_a}{R_u} w_5 w_{6} \dot{T},   &\dot{w}_6 & =  -2T w_{6}^2, \\
0  & = w_7 -  w_1 w_3 w_5.
\end{align*}	
The temperature $T = \frac{p \xi}{R(Y_l)}$ can be obtained from the states  $\xi,p, Y_l$ via the ideal gas relationship in Eq.~\eqref{eq:idealGas}.

\section*{Appendix B: Equations for pressure and chemical species} \label{app:pressure}
Here, we give the complete governing equation for the species molar concentrations $c_l$ and pressure $p$. Recall from Section~\ref{sec:transform} the notation used to denote the conservative variables $g_i$: $g_1 = \rho, \ g_2 = \rho v_x, \ g_3 = \rho v_y, \ g_4 = \rho e, \ g_5 = \rho Y_1, \ g_6 = \rho Y_2, \ g_7 = \rho Y_3, \ g_8 = \rho Y_4$.\\

\noindent \textbf{Species molar concentrations $c_l$:}  For the species molar concentration dynamics, we use the relationship $c_l = \frac{\rho Y_l}{M_l}$ from Eq.~\eqref{eq:moletomass}, where the constants $M_1, \ldots, M_4$ are molar masses. We obtain for $l = 1,2,\ldots,n_\text{sp}$:
\begin{align*}
\frac{\partial c_l}{\partial t}
& = \frac{1}{M_l} \frac{\partial \rho Y_l}{\partial t}\\
& =  \frac{1}{M_l}  \left ( \dot{\omega}_l +  \nabla \cdot \left (- v_x \rho Y_l\vec{i} -  v_y \rho  Y_l \vec{j} + \vec{j}_{l}^m   \right ) \right )\\
& =  \frac{1}{M_l}  \left ( M_l \dot{c}_l^\text{reaction} -  M_l \left (  \frac{\partial v_x c_l }{\partial x} + \frac{\partial v_y c_l }{\partial y} \right )    + \nabla \cdot  \vec{j}_{l}^m  \right )\\
& =  \dot{c}_l^\text{reaction} - \left (  \frac{\partial v_x c_l }{\partial x} + \frac{\partial v_y c_l }{\partial y} +  \frac{1}{M_l} \nabla \cdot  \vec{j}_{l}^m  \right ).
\end{align*}
Note that the chemical source terms $\dot{c}_l^\text{reaction}$ were given in Appendix~A, and from Eq.~\eqref{eq:chemistry} we see that the $\dot{c}_l$ are cubic in the auxiliary lifted states $w_1, \ldots, w_6$.
The divergence term is
\begin{equation*}
\nabla \cdot  \vec{j}_{l}^m
=  D_l \left (  \frac{\partial }{\partial x} \left (\rho \frac{\partial Y_l}{\partial x} \right ) + \frac{\partial }{\partial y} \left  (\rho \frac{\partial Y_l}{\partial y} \right ) \right ),
\end{equation*}
and since $Y_l = M_l c_l \xi$, we have that
\begin{equation*}
\rho \frac{\partial Y_l }{\partial x} = \rho M_l \frac{\partial c_l \xi }{\partial x} = M_l \left (\frac{\partial c_l }{\partial x} + \rho c_l \frac{\partial \xi }{\partial x}  \right ).
\end{equation*}
Overall, we have that
\begin{equation*}
\frac{\partial c_l}{\partial t} = \dot{c}_l^\text{reaction} - \left (  \frac{\partial v_x c_l }{\partial x} + \frac{\partial v_y c_l }{\partial y} + \frac{\partial c_l }{\partial x} + \rho c_l \frac{\partial \xi }{\partial x} + \frac{\partial c_l }{\partial y} + \rho c_l \frac{\partial \xi }{\partial y}  \right ),
\end{equation*}
which is quadratic in the learning variables $v_x, v_y, c_l$ with the exception of the terms $\rho c_l \frac{\partial \xi }{\partial x}$ and $\rho c_l \frac{\partial \xi }{\partial y}$. We note that if $\rho$ were included as a lifted variable (in addition to $\xi$), these terms would become quadratic in the lifted state.\\

\noindent \textbf{Pressure $p$:}
We start with the energy equation~\eqref{eq:e}. By multiplying with density $\rho$, we have $\rho e = \rho h^0 -p$, so from the conservation equation \eqref{eq:ns_conservative} for $\rho e$  we obtain
\begin{equation*}
\frac{\partial (\rho h^0-p)}{\partial t} + \frac{\partial \rho v_x h^0}{\partial x} + \frac{\partial \rho v_y h^0}{\partial y} + \frac{\partial }{\partial x} (v_x \tau_{xx} + v_y \tau_{yx} -j_x^q ) + \frac{\partial }{\partial y} (v_x \tau_{xy} + v_y \tau_{yy} -j_y^q ) =0.
\end{equation*}
This directly gives an equation for the time evolution of pressure:
\begin{equation*}
\frac{\partial p}{\partial t} = \frac{\partial \rho h^0}{ \partial t} + \frac{\partial \rho v_x h^0}{\partial x} + \frac{\partial \rho v_y h^0}{\partial y} + \frac{\partial }{\partial x} (v_x \tau_{xx} + v_y \tau_{yx} -j_x^q ) + \frac{\partial }{\partial y} (v_x \tau_{xy} + v_y \tau_{yy} -j_y^q ).
\end{equation*}
Moreover, per definition of $h^0$ in Eq.~\eqref{eq:e} and with $c_l = \frac{\rho Y_l}{M_l}$ we have that
\begin{equation*}
\frac{\partial \rho h^0}{\partial t}
= \sum_{i=1}^{n_\text{sp}} \frac{\partial \rho h_l Y_l}{\partial t} + \frac{ \partial \rho \frac{1}{2} \left (v_x^2 + v_y^2 \right )}{\partial t}
= \sum_{i=1}^{n_\text{sp}} M_l \frac{\partial h_l c_l}{\partial t}+ \frac{1}{2} \left (v_x^2 + v_y^2 \right ) \frac{\partial \rho}{\partial t} + \rho \frac{\partial (v_x + v_y)}{\partial t}.
\end{equation*}
Overall, we have that
\begin{align*}
\frac{\partial p}{\partial t} =&   \sum_{i=1}^{n_\text{sp}} M_l \left ( h_l  \frac{\partial c_l}{\partial t} + c_l \frac{\partial h_l}{\partial t} \right ) + \frac{1}{2} \left (v_x^2 + v_y^2 \right ) \frac{\partial \rho}{\partial t} + \rho \frac{\partial (v_x + v_y)}{\partial t} + \frac{\partial \rho v_x h^0}{\partial x} + \frac{\partial \rho v_y h^0}{\partial y}\\
 + & \frac{\partial }{\partial x} (v_x \tau_{xx} + v_y \tau_{yx} -j_x^q ) + \frac{\partial }{\partial y} (v_x \tau_{xy} + v_y \tau_{yy} -j_y^q ).
\end{align*}
This equation remains nonlinear in our chosen learning variables $\vec{q}_L$. In particular, the enthalpies $h_l = h_l(T)$ and their time derivatives are nonlinear functions of temperature. The other terms show some polynomial structure; for example, in Section~\ref{sec:transform} we showed that $\frac{\partial (v_x + v_y)}{\partial t}$ is quadratic in the learning state variables $p, v_x, v_y, \xi$. However, to write this equation exactly in a polynomial form would require introducing a large number of auxiliary variables along with their corresponding dynamics. We have instead chosen to introduce an approximation by learning a ROM in the variables $\vec{q}_L$ with quadratic form.

\section*{Funding Sources}
This work has been supported in part by the Air Force Center of Excellence on Multi-Fidelity Modeling of Rocket Combustor Dynamics award FA9550-17-1-0195, and the Air Force Office of Scientific Research (AFOSR) MURI on managing multiple information sources of multi-physics systems award numbers FA9550-15-1-0038 and FA9550-18-1-0023.

\bibliography{sample,sample_elizabeth}
\bibliographystyle{aiaa}
\end{document}